\RequirePackage{fix-cm}
\documentclass[11pt,a4paper]{article}  

\usepackage[latin9]{inputenc}
\usepackage{float}
\usepackage{mathtools}
\usepackage{amsmath}
\usepackage{graphicx}
\usepackage{xcolor}
\usepackage{ulem}
\usepackage{bbm}

\usepackage{amssymb}
\usepackage{algorithm}
\usepackage[noend]{algpseudocode}
\usepackage{epsfig}
\usepackage{pgfplots}
\pgfplotsset{/pgf/number format/use comma,compat=newest}
\usepackage{subfigure}
\usepackage{graphics}
\usepackage[english]{babel}

\newcommand{\Kcap}{\widehat{K}}
\newcommand{\xcap}{\hat{\bf x}}
\newcommand{\x}{{\bf x}}
\newcommand{\T}{{\mathcal T}}

\newcommand{\F}{{\mathcal F}}
\newcommand{\ra}{{\bf r}_{1, K}}

\newcommand{\rc}{{\bf r}_{d, K}}
\newcommand{\ri}{{\bf r}_{i, K}}

\newcommand{\la}{\lambda_{1, K}}

\newcommand{\lc}{\lambda_{d, K}}
\newcommand{\li}{\lambda_{i, K}}

\newcommand{\M}{{\mathcal M}}

\newcommand{\gi}{{\bf g}_i}

\newcommand{\comment}[1]{ }

\newtheorem{propo}{Proposition}[section]

\floatstyle{ruled}
\newfloat{algorithm}{tbp}{loa}
\providecommand{\algorithmname}{Algorithm}
\floatname{algorithm}{\protect\algorithmname}

\begin{document}

\title{
Enhancing level set-based topology optimization with anisotropic graded meshes}

\author{Davide Cortellessa$^1$, Nicola Ferro$^2$, Simona Perotto$^2$,\\ 
Stefano Micheletti$^2$}
\maketitle

\begin{center}
{\small
$^1$
Dipartimento di Matematica\\
Politecnico di Milano\\
Piazza L. da Vinci, 32, I-20133 Milano, Italy\\
{\tt davide.cortellessa@mail.polimi.it}
\\[3mm]
$^2$
MOX -- 
Dipartimento di Matematica\\
Politecnico di Milano\\
Piazza L. da Vinci, 32, I-20133 Milano, Italy\\
{\tt \{nicola.ferro, simona.perotto, stefano.micheletti\}@polimi.it}
}
\end{center}
\date{}
\maketitle

\begin{abstract}
We propose a new algorithm for the design of topologically optimized light\-weight structures, under a minimum compliance requirement. The new process enhances a standard level set formulation in terms of computational efficiency, thanks to the employment of a strategic computational mesh. We pursue a twofold goal, i.e., to deliver a final layout characterized by a smooth contour and reliable mechanical properties.
The smoothness of the optimized structure is ensured by the employment of an anisotropic adapted mesh, which sharply captures the material/void interface. A robust mechanical performance is guaranteed by a uniform tessellation of the internal part of the optimized configuration. A thorough numerical investigation corroborates the effectiveness of the proposed algorithm as a reliable and computationally affordable design tool, both in two- and three-dimensional contexts.
\end{abstract}

\section{Introduction}

Topology optimization is of utmost interest in different branches of industrial design and engineering. The main objective is to devise how to place material in a given design domain with the aim of obtaining the best performance, according to specific criteria.
This idea, firstly proposed for mechanical applications, has been extended to a variety of fields, such as biomedical, space, automotive, fluids, acoustics, electromagnetics, optics, architecture, design of new materials (see, e.g., \cite{sigmund2013, alaimo2017,
collet2017, zargham2016, Ferro2021, huang2012}).
\\
Independently of the specific field of application, the mathematical framework is represented by a constrained optimization problem, so that a quantity of interest is minimized under assigned constraints. For instance, in this paper, we focus on the design of lightweight and stiff structures by minimizing the compliance with a constraint on the mass, in a linear elasticity regime. 

Several methods are available in the literature to address the formulation of topology optimization. 
Among these, we mention the density-based approaches~\cite{bendsoe1995, sigmund2004, rozvany}, the level set methods~\cite{allaire2004structural, yulin2004, wang2003}, topological derivative procedures~\cite{sokolowski2009topological}, phase field techniques~\cite{bourdinchambolle03, dede2012isogeometric}, evolutionary algorithms~\cite{xie1997basic}, homogenization~\cite{allaire2004, bendsoe1988}, performance-based optimization~\cite{liang2005}. Comprehensive reviews and comparisons of all these approaches are covered in~\cite{sigmund2004,sigmund2013}.
\\
In this paper, we are primarily concerned with the level set method. The so-called level set function, whose zero-contour identifies the boundary of the material layout, is evolved in order to target the objective functional and the imposed constraints. With this aim, it is standard to resort to a diffusion equation in a pseudo-time setting, relying on a topological derivative and a Laplacian smoothing term~\cite{yamada2010topology}. Among the most recent works where the level-set method is applied to topology optimization, we cite~\cite{cohen2022,andrade2022,oka2022,cui2022,zhou2022,dapogny2021,nardoni2022}.

The new contribution of this paper is represented by the enhancement of the standard level set formulation, by means of a computationally strategic selection of the design domain discretization, in a finite element setting. 
With this aim, we adopt an advanced metric-based mesh adaptation technique for anisotropic triangular~\cite{formaggia2001new,wccm2002} and tetrahedral elements~\cite{micheletti2010recovery,farrell2011anisotropic}, driven by an a posteriori recovery-based error estimator~\cite{zienkiewicz_zhu2,zienkiewicz_zhu4}.
In addition, in order to guarantee a reliable mechanical analysis, we modify the fully anisotropic metric into a graded spacing, where stretched elements are preserved to describe the material boundary, whereas an isotropic tessellation is used in the interior portion of the structure~\cite{ferro2020c}. 
We name the newly proposed algorithm LEVITY (LEVel set with mesh adaptivITY). In a similar spirit, in~\cite{rodenas2022}, the authors carry out topology optimization through $h$-refinement and coarsening, but restricted to the context of isotropic Cartesian meshes.

This paper aims at verifying the good properties of LEVITY that can be thus listed:
\begin{itemize}
    \item automation, that is guaranteed for free by the adopted metric-based mesh adaptation process~\cite{farrell2011anisotropic};
    \item cost-effectiveness, that is supported by the well-established advantages ensured by adapted (anisotropic) meshes when compared with fixed grids (see, e.g.,~\cite{roux2013,soli2019,fortin2002,cangiani22,perotto2022}), and by the computational cheapness of the recovery-based error estimators here exploited~\cite{AO};
    \item limited post-processing of the optimized structure before manufacturing, thanks to the smoothness of the optimized layout boundary provided by the allocation of highly stretched elements;
    \item mechanical reliability, guaranteed by the employment of isotropic elements inside the structure, thus preserving the approximation of the compliance by any bias~\cite{ferro2020c}.  
\end{itemize}

Moreover, a numerical comparison between LEVITY and the standard level set approach for topology optimization highlights that the new algorithm essentially preserves the good properties of the original one, such as the independence of the final layout with respect to both the initial topology and computational grid. 

In short, LEVITY can be classified as a mechanically reliable and computationally convenient design tool able to provide optimized structures characterized by very clear-cut contours, at a limited computational effort. This allows us to deal with challenging $3$D configurations which, in general, are not efficiently affordable with standard topology optimization tools.

The paper is organized as follows.
Section~\ref{sec:levelset_generic} introduces the minimum compliance problem together with the standard level set approach for topology optimization. Section $3$ formalizes the LEVITY algorithm, after providing the metric-based anisotropic mesh adaptation basics. In Section~\ref{numerical_sec}, we perform an extensive numerical assessment of LEVITY algorithm on $2$D case studies, which are benchmark in the literature. Successively, we challenge LEVITY on realistic $3$D configurations in Section~\ref{sec:3d}. Finally, some conclusions and future perspectives are drawn in the last section.

\section{The minimum compliance design problem: a level set approach}\label{sec:levelset_generic}

We consider a generic topology optimization framework which relies on a function $\chi \in X$, that is used to label the void ($\chi({\bf x}) = 0$, with ${\bf x} \in \Omega$) or the material ($\chi({\bf x}) = 1$, with ${\bf x} \in \Omega$) to be alternated inside the design domain $\Omega \subset \mathbb{R}^d$, with $d = 2, 3$. In particular, we assume as a reference setting the optimization problem
\begin{equation}\label{eq:topopt_chi}
\min_{\chi \in X} F(u(\chi), \chi)  :
\left\{
\begin{array}{l}
\mathcal{S}(u(\chi), \chi; v) = 0 \quad \forall v \in U\\[2mm]
\mathcal{C}_{i,m} \le \mathcal{C}_i(u(\chi), \chi) \le \mathcal{C}_{i,M} \quad i = 1, ..., N_{\mathcal{C}}\\[3mm]
\chi \in [\chi_{min}, 1],
\end{array}
\right.
\end{equation}
where $u = u(\chi) \in U$ denotes the state function; $F$ represents the goal functional, which a priori may depend on both the design, $\chi$, and the state, $u$, variables, with $U$ and $X$ function spaces to be properly selected. The optimization process is constrained by the state equation, $\mathcal{S}(u(\chi), \chi; v) = 0$, which coincides with the weak formulation of the partial differential equation (PDE) model characterizing the physics of interest; by $N_{\mathcal{C}}$ box inequalities, which involve the physical quantities $\mathcal{C}_i$ strictly related to the configuration at hand, with $\mathcal{C}_{i,m}$ and $\mathcal{C}_{i,M}$ the corresponding lower and upper bounds, respectively; by the admissible range of values for the design function $\chi$, with $\chi_{min} \ge 0$ arbitrarily small.

The topology process in \eqref{eq:topopt_chi} is fully general. A suitable choice for the goal functional, for the state equation, and the constraints can cast the optimization into different scenarios. In particular, the focus of this paper is on the minimization of the compliance of a structure which is loaded by the external traction $\boldsymbol{t}$ along the portion $\Gamma_{t}$ of the domain boundary $\partial\Omega$, under a linear elasticity regime \cite{gould}. This context provides a well-established benchmark case study~\cite{bendsoe,bendsoe1995,sigmund2004}. Thus, the goal functional in \eqref{eq:topopt_chi} coincides with the compliance
\begin{equation}\label{compliance}
F({\bf u}(\chi), \chi) = \int_{\Gamma_{t}}\boldsymbol{t}\cdot{\bf u}(\chi)\,d\gamma,
\end{equation}
with ${\bf u} = [u_1, \ldots, u_d]^T$ the displacement induced by the imposed load; the state equation is framed in a suitable subset $U$ of the Sobolev space $H^1(\Omega)$, which accounts for possible essential boundary conditions \cite{ern}, being
\begin{equation}
{\mathcal S} ({\bf u}(\chi), \chi; {\bf v}) = a_\chi({\bf u}(\chi),{\bf v}) - l({\bf v}) = 0,\label{eq:weak_form}
\end{equation}
with
\begin{equation}\label{forms}
a_\chi({\bf u} (\chi),{\bf v}) = \displaystyle \int_{\Omega} \chi \sigma({\bf u} (\chi))\vcentcolon\varepsilon({\bf v})\,d\Omega, \quad l({\bf v})=\int_{\Gamma_{t}}\boldsymbol{t}\cdot{\bf v}\,d\gamma.
\end{equation}
In the linear elasticity regime under consideration, the stress tensor is
$\sigma({\bf u}(\chi))=2\mu\varepsilon({\bf u}(\chi))+\lambda\textrm{tr}(\varepsilon({\bf u} (\chi)))I$, where $\varepsilon({\bf u} (\chi))={(\nabla{\bf u}(\chi)+\nabla{\bf u}(\chi)^{T})}/{2}$ denotes the strain tensor, $\textrm{tr}(\cdot)$ the trace operator,
$I \in \mathbb{R}^{d \times d}$ the identity matrix, $\lambda={E\nu}/[(1+\nu)(1-2\nu)]$ and $\mu=E/[2(1+\nu)]$ the Lam\'e coefficients, $E$ being the Young modulus and $\nu$ the Poisson ratio.
\\
In particular, we highlight that in \eqref{forms} the variable $\chi$ modifies the standard linear elasticity equation in order to drive the design of the topologically optimized structure, here denoted by $\Sigma$.
Moreover, the linear functional in \eqref{forms} does coincide with the compliance in \eqref{compliance}, for ${\bf v} = {\bf u}(\chi)$.
\\
Concerning the box constraints in \eqref{eq:topopt_chi}, we resort to a control on the maximum allowed material quantity, by setting $N_\mathcal{C} = 1$ and by identifying
$$
\mathcal{C}_1({\bf u}(\chi), \chi) = \mathcal{C}(\chi) = \displaystyle \int_\Omega \chi\, d\Omega, \quad \mathcal{C}_{1,m} = 0, \quad \mathcal{C}_{1,M} = \alpha V_0,
$$
where $\alpha \in (0,1]$ is the prescribed volume fraction with respect to the initial volume of the structure, $V_{0}=\int_{\Omega}1 d\Omega$.

Thus, the minimum compliance problem reads as:
\\find $\chi \in L^\infty(\Omega)$ such that:

\begin{equation}\label{eq:min_comp_chi}
\min_{\chi\in  L^\infty(\Omega)}l({\bf u}(\chi)) :
\left\{
\begin{array}{l}
a_\chi({\bf u}(\chi),{\bf v})-l({\bf v}) = 0 \quad \forall{\bf v}\in U
\\[3mm]
\mathcal{C}(\chi) \leq\alpha V_{0}
\\[3mm]
\chi \in [\chi_{min}, 1].
\end{array}
\right.
\end{equation}
Notice that the lower bound for $\mathcal{C}(\chi)$ is automatically satisfied due to \eqref{eq:min_comp_chi}$_3$.\\
In the following, the dependence of ${\bf u}$ on the design variable $\chi$ will be dropped to simplify the notation. 

\subsection{The level set formulation}\label{LS_sec}

The choice for the design function $\chi$ in \eqref{eq:topopt_chi} is not unique. According to a level set approach, the boundary, $\partial \Sigma$, of the structure, $\Sigma\subset \Omega \subset \mathbb{R}^d$, to be designed is described by the zero-isocontour of a $(d+1)$-dimensional surface $\Gamma: \Omega \rightarrow \mathbb{R}^{d+1}$ so that $\Gamma({\bf x}) =  ({\bf x}, \varphi({\bf x}))$, with $\varphi$ the level set function. In particular, we select $\varphi({\bf x}): \Omega \rightarrow [-1, 1]$, to be
\begin{equation}
\begin{cases}
0 < \varphi \le 1 & \textrm{for}\,{\bf x}\in\Sigma\\
\varphi = 0 & \textrm{for}\,{\bf x}\in\partial\Sigma\\
-1 \le \varphi < 0 & \textrm{for}\,{\bf x}\in \Omega\backslash \overline\Sigma.
\end{cases}\label{ls:4}
\end{equation}
The optimal layout corresponds to the portion in $\Omega$ where the level set function takes  non-negative values. As a consequence, function $\chi$ in \eqref{eq:topopt_chi} can be identified with
\begin{equation}\label{CHI}
\chi_\varphi = 
\left\{
\begin{array}{ll}
    1          &   \varphi \ge 0\\
    \chi_{min} &   \varphi < 0,
\end{array}
\right.
\end{equation}
Smoothness assumptions are made both on functions $\varphi$ and $\chi_{\varphi}$, namely $\varphi \in C^0(\overline{\Omega})$, while $\chi_\varphi \in L^2(\Omega)$.
\\
Thus, the minimum compliance problem in \eqref{eq:min_comp_chi} with the level set approach is formulated as:
\\
find $\varphi\in H^{1}(\Omega; [-1, 1])$ such that:
\begin{equation}\label{level_set}
\min_{\varphi\in H^{1}(\Omega; [-1, 1])}\,l({\bf u}) \ :
\left\{
\begin{array}{l}
a_{\chi_\varphi}({\bf u}, {\bf v}) - l({\bf v}) = 0 \quad \forall {\bf v}\in U\\[3mm]
\displaystyle \mathcal{C}(\chi_{\varphi}) \le \alpha V_{0},
\end{array}
\right.
\end{equation}
where $a_{\chi_\varphi}(\cdot, \cdot)$ is defined as in \eqref{forms} with $\chi = \chi_\varphi$, and where the last constraint in \eqref{eq:min_comp_chi} directly follows from the definition of $\chi_\varphi$.

According to a level set approach, the identification of the optimized solution to  problem \eqref{level_set} relies on a time-dependent process, which evolves an initial contour, $\varphi^0$, towards the final layout boundary $\partial \Sigma$. Following, for instance, \cite{yamada2010topology,otomori2014}, such evolution is governed by the diffusive process 
\begin{equation}
\begin{cases}
{\displaystyle \frac{\partial\varphi}{\partial t}}=\kappa \, d_{t}\overline{F}+\tau\Delta\varphi & \textrm{in } \Omega,\,\,\, t>0\\[3mm]
{\displaystyle \frac{\partial\varphi}{\partial n}=0} & \textrm{on }\partial \Omega,\,\,\, t>0\\[3mm]
{\displaystyle \varphi=\varphi^{0}} & \textrm{in } \Omega,\,\,\, t = 0,
\end{cases}\label{evolution}
\end{equation}
with $\tau > 0$ a parameter tuning the diffusivity of the level set evolution;
$$
\overline{F} = \overline{F}({\bf u}, {\bf w}, \theta; \varphi) = l({\bf u}) - \big[a_{\chi_{\varphi}}({\bf u}, {\bf w}) - l({\bf w})\big] + \theta G(\varphi),
$$
the Lagrangian functional that commutes the constrained optimization in \eqref{level_set} into an unconstrained procedure, by introducing the Lagrangian multipliers ${\bf w} \in U$ and $
\theta \in \mathbb{R}^+$, being
$$
G(\varphi) = \mathcal{C}(\chi_\varphi) - \alpha V_0 \le 0;
$$
$d_{t}\overline{F}$ the topological derivative of the Lagrangian functional $\overline{F}$~\cite{yamada2010topology,sokolowski2009topological,kao2007,novotny2003}; $\kappa > 0$ a normalization factor.
In particular, the minimum compliance problem in \eqref{eq:min_comp_chi} leads to identify the adjoint with the state problem, i.e., ${\bf w} = {\bf u}$ (see, \cite{yamada2010topology,otomori2014} for all the details).

We remark that the diffusive contribution $\tau\Delta\varphi$ ensures the smoothness of the level set function, acting as a perimeter control. Thus, by properly tuning parameter $\tau$, we can limit or promote the generation of complex geometric features~\cite{yamada2010topology,otomori2014}.
\\
Finally, the differential problem in \eqref{evolution} is completed by homogeneous Neumann boundary conditions for simplicity of implementation. 

\subsection{The level set discrete formulation}

We numerically deal with problem \eqref{evolution} by adopting a standard continuous finite element discretization for the spatial dependence, combined with a backward Euler scheme to approximate the time evolution~\cite{ern}.
In addition, in order to compute the Lagrangian functional $\overline{F}$ we have to approximate the state equation in \eqref{level_set}.

To this aim, we discretize domain $\Omega$ with a family of conforming tessellations, $\{\T_h\}$, characterized by triangular/tetrahedral elements $K$. Analogously, the considered time window is partitioned by the discrete instants $\{t^{\tt k}\}$ into uniform time intervals of length $\Delta t$.
\\
The approximation of the state equation in \eqref{level_set} and of problem \eqref{evolution} is performed in the discrete spaces $U_h = [X_h^1]^d \cap U$, and $\Phi_h = X_h^1$, which leads to replace ${\bf u}$ with ${\bf u}_h = [u_{1,h}, \ldots, u_{d,h}]^T \in U_h$, $\varphi$ with $\varphi_h \in \Phi_h$, being
$$
X_h^1 = \{f \in C^0(\overline{\Omega}) \, \mbox{ s.t. } \, f|_K \in \mathbb{P}^1_K, \, \forall K \in \T_h\}, 
$$
the space of the affine finite elements, and $\mathbb{P}^1_K$ the polynomials of degree one with real coefficients.
Space $X_h^1$ is also adopted to approximate the characteristic function $\chi_\varphi$ in \eqref{CHI} by $\chi_{\varphi,h}$. The employment of a unique discrete space to approximate all the functions involved in the level set problem relieves us from any projection step among different spaces.

In Algorithm~\ref{level_set_algo}, we formalize the level set approach implementing the discrete minimum compliance design problem.

\begin{algorithm}[H]
	\caption{Minimum compliance topology optimization with level set}\label{level_set_algo}
	{\bf Input}:  {\tt CTOL, kmax}, $\varphi_{h}^{\tt 0}$, $\T_h^{\tt 0}$, $\Delta t$, $\alpha$, $V_0$, $\chi_{min}$, $\tau$
	\begin{algorithmic}[1]
		\State Set: ${\tt k} = {\tt 0}$, errComp = 1+{\tt CTOL} \vspace{1mm}
			\While {errComp $>$ {\tt CTOL} \& ${\tt k} <$ {\tt kmax}} \vspace{2mm}
            \State $\chi_{\varphi,h}^{\tt k}$ = {\tt characteristic}($\varphi_h^{\tt k}$, $\T_h^{\tt 0}$, $\chi_{min}$);\label{char_al} \vspace{2mm}
			\State ${\bf u}_h^{\tt k+1}$ = {\tt solveState}($\chi_{\varphi,h}^{\tt k}$, ${\bf v}_h$);\label{state_al}\vspace{1mm}
			\State [$d_{t}\overline{F}^{\tt k+1}$, $\kappa$] = {\tt topologicalDerivative}(${\bf u}_h^{\tt k+1}$, $\alpha$, $V_0$);\label{topo_al}\vspace{1mm}
			\State $\tilde{\varphi}_h^{\tt k+1}$ = {\tt evolveLevelSet}($\varphi_h^{\tt k}$, $d_{t}\overline{F}^{\tt k+1}$, $\kappa$, $\tau$, $\Delta t$);\label{evolve_al}\vspace{2mm}
			\State $\varphi_h^{\tt k+1}$ = {\tt threshold}($\tilde{\varphi}_h^{\tt k+1}$);\label{thr_al}\vspace{2mm}
			\State errComp = $|l({\bf u}_h^{\tt k+1})-l({\bf u}_h^{\tt k})|/|l({\bf u}_h^{\tt k})|$;\vspace{2mm}
			\State ${\tt k} = {\tt k+1}$; 
			\EndWhile
			\State {\bf end while} \vspace{2mm}
			\State $\overline{\Sigma}_h$ = {\tt extract}($\varphi_h^{\tt k}$);\label{extract_al} \vspace{2mm}
			\end{algorithmic}
			{\bf Output}: $\overline{\Sigma}_h$
\end{algorithm}

\noindent
The input values for the algorithm are $\tt CTOL$ and $\tt kmax$ to rule the {\bf while} loop; the initial level set function $\varphi_{h}^{\tt 0}$ defined on the mesh $\T_h^{\tt 0}$; the time step $\Delta t$; the maximum prescribed volume fraction, $\alpha$, with respect to the full-material configuration volume, $V_0$; the lower bound, $\chi_{min}$, for function $\chi_{\varphi, h}$; the coefficient $\tau$ driving the diffusive process in \eqref{evolution}.

The algorithm essentially consists of five steps performed at each time $t^{\tt k}$. In particular, there exists a correspondence between the index $\tt k$ of the {\bf while} loop and the instant $t^{\tt k}$, such that ${\tt k} = {\tt k} + 1$ amounts to $t^{\tt k+1} = t^{\tt k} + \Delta t$. \\
For each ${\tt k}$, we first identify the characteristic function $\chi_{\varphi,h}^{\tt k}$ by computing the zero-isocontour of $\varphi_h^{\tt k}$ (line \ref{char_al}). Then, we solve the state equation in \eqref{level_set} to update the discrete displacement (line \ref{state_al}). The new displacement and the quantities characterizing the inequality constraint in \eqref{level_set} are provided as the inputs to compute the topological derivative $d_t \overline{F}$ and the factor $\kappa$ (line \ref{topo_al}). 
\\
Successively, the level set function $\varphi_h^{\tt k}$ is evolved according to the model in \eqref{evolution} into $\tilde{\varphi}_h^{\tt k+ 1}$ (line \ref{evolve_al}), which is then thresholded (line \ref{thr_al}) by the rule
$$
\textrm{if }|\tilde{\varphi}_h^{\tt k+ 1}|>1\textrm{ then } {\varphi}_h^{\tt k+ 1}=\textrm{sign}(\tilde{\varphi}_h^{\tt k+ 1}).
$$
The five-step procedure just detailed is constrained by a control on the relative accuracy, ${\tt CTOL}$, on the compliance combined with a maximum number of time steps, ${\tt kmax}$. 

Algorithm~\ref{level_set_algo} returns the final layout $\overline{\Sigma}_h = \{{\bf x} \in \Omega \, \mbox{ s.t. } \, \varphi_h^{\tt k}({\bf x}) \ge 0\}$, coinciding with the minimum compliance structure (line \ref{extract_al}).

\section{Level set enriched by mesh adaptation}
This section provides an advanced algorithm for problem \eqref{level_set} where a standard level set approach is enhanced by discretizing the PDE problems in \eqref{level_set} and \eqref{evolution} on customized computational grids able to follow the progressive design of the optimized structure.

\subsection{A metric-based anisotropic mesh adaptation}\label{aniso_sec}
Anisotropic mesh adaptation proved to be the optimal tool for modeling phenomena characterized by preferential directions, e.g., in the presence of boundary or internal layers, shocks, wakes, in diverse application fields~\cite{roux2013,soli2019,micheletti2010recovery,micheletti2008output,farrell2011anisotropic,fortin2002,cangiani22}.
\\
An anisotropic mesh adaptation procedure consists of an iterative modification of the numerical grid by optimally adjusting the size, the orientation and the shape of the elements, in order to sharply track the occurring directionalities. In contrast to an isotropic mesh adaptation, which only prescribes the optimal size of the elements while keeping the shape fixed, the anisotropic approach guarantees more versatility and a higher efficiency (for instance, by reducing the number of elements for a user-defined accuracy).

In the sequel, the anisotropic mesh generation is carried out in the well-established setting proposed in the seminal works~\cite{formaggia2001new,micheletti2010recovery,farrell2011anisotropic}. With reference to a $d$-dimensional context, we recover the geometric description of a generic element $K \in \mathcal{T}_h$, out of the spectral properties of the standard affine map, $T_K: \Kcap \rightarrow K$, such that
$$
\x=T_K(\xcap) = M_K \xcap +{\bf t}_K,
$$ 
with $ \x \in K$ and $\xcap \in \Kcap$, $\Kcap$ being the reference element inscribed into the unit $d$-sphere. In particular, $M_K \in \mathbb{R}^{d \times d}$ is the Jacobian of map $T_K$ and is responsible for the transformation of the unit $d$-sphere into a $d$-ellipsoid circumscribing $K$, whereas ${\bf t}_K\in \mathbb{R}^d$ represents a rigid translation.
The geometric description of $K$ is obtained by considering:
\begin{enumerate}
\item[i)] the polar decomposition $M_K = B_K Z_K$ of the Jacobian, with $B_K \in \mathbb{R}^{d\times d}$ a symmetric positive definite deformation matrix and $Z_K\in \mathbb{R}^{d\times d}$ a rotation;
\item[ii)] the standard spectral decomposition of matrix $B_K$, so that $B_K=R_K^T \Lambda_K R_K$, with $R_K^T=[\ra, \ldots, \rc]$ and $\Lambda_K={\rm diag}(\la, \ldots$ $, \lc)$, the matrices of the eigenvectors and the eigenvalues of $B_K$, with $\la \ge \ldots \ge \lc > 0$.
\end{enumerate}
Matrices $R_K^T$ and $\Lambda_K$ completely characterize the element $K$. In particular, the eigenvectors $\ra, \ldots, \rc$ are aligned with the directions of the semi-axes of the $d$-ellipsoid circumscribed to $K$, while the eigenvalues $\la, \ldots, \lc$ measure the length of such semi-axes (we refer to~\cite{formaggia2001new,farrell2011anisotropic} for more details).

To evaluate the anisotropy of the element $K$ with respect to the reference isotropic case, we adopt the quantities
\begin{equation*}
s_{i, K}=
\left(
\lambda_{i, K}^{2(d - 1)/d}
\right)
\bigg(
{\displaystyle \prod_{j = 1, \, j \neq i}^d \lambda_{j, K}}\bigg)^{-2/d}
\quad i = 1, \ldots, d,
\end{equation*}
which are referred to as aspect ratios. In particular, $s_{i, K}$'s are identically equal to $1$ for an isotropic element.

\subsubsection{The error estimator}\label{subsubsect_212}
An anisotropic mesh adaptation procedure can be driven by imposed criteria, which are possibly related to the discrete solution under investigation. Common strategies rely on suitable estimators for the discretization error in order to identify the portions of the domain where the tessellation needs to be refined, coarsened or deformed in order to capture the directionalities of the solution at hand.

In this paper, we resort to a recovery-based analysis which consists of two steps, i.e., the computation of a so-called recovered gradient and the definition
of the associated error estimator~\cite{zienkiewicz_zhu2,zienkiewicz_zhu3,zienkiewicz_zhu4}.
In more detail, we adopt the anisotropic counterpart of the estimator in~\cite{zienkiewicz_zhu4}, proposed for the first time in~\cite{enumath09}.

As a first step, we define the recovered gradient, $P(\nabla \cdot)$, by means of an area-weighted average of the discrete gradient across the patch of elements, $\Delta_K = \{T\in \T_h: T\cap K \neq \emptyset\}$ associated with $K$,
\begin{equation}\label{recovered_gradient}
P(\nabla w_h)({\bf x})=\frac{1}{|\Delta_K|} \sum_{T \in \Delta_K}
|T|\, \nabla w_h|_T \quad {\bf x} \in K,
\end{equation}
with $w_h$ a finite element approximation of the generic function $w \in H^1(\Omega)$, and $|\varpi|$ the measure of the $d$-dimensional set $\varpi$.
Notice that, following~\cite{farrell2011anisotropic,micheletti2010recovery}, we select a piecewise constant recovered gradient to simplify the computations, by involving a sufficiently large number of mesh elements in the average step.

Successively, we introduce an anisotropic a posteriori error estimator based on the recovered gradient  \eqref{recovered_gradient}, following~\cite{enumath09,micheletti2010recovery}. 
The global estimator, $\eta$, can be characterized in terms of the elementwise contributions, $\eta_K$ by
\begin{equation}\label{globestim}
\eta^2 =\sum_{K\in \T_h} \eta_K^2,
\end{equation}
with 
\begin{equation}\label{locestim}
\eta_K^2 =
\bigg({\displaystyle \prod_{j = 1}^d \lambda_{j, K}}\bigg)^{-2/d}
\
\sum_{i=1}^{d}
\li^2\, \Big(\ri^T\, G_{\Delta_K}\big( E_\nabla \big)\ri\,\Big).
\end{equation}
Here, $E_{\nabla}=\big[P(\nabla w_h) - \nabla w_h\big]$ denotes the recovered error on the gradient, while
$G_{\Delta_K}(\cdot)\in \mathbb{R}^{d\times d}$ is a symmetric positive semidefinite matrix with entries
\begin{equation}\label{micheletti_G} 
[G_{\Delta_K}({\bf q})]_{i,j} = \sum_{T\in \Delta_K} \int_T q_i\,q_j \, dT \quad \mbox{with } i,j=1,\ldots,d,
\end{equation} 
for any vector-valued function ${\bf q}=(q_1,\ldots,q_d)^T \in [L^2(\Omega)]^d$. The local estimator \eqref{locestim} coincides with the anisotropic counterpart of the $L^2(K)$-norm of the recovered error $E_\nabla$, 
scaled by factor $({\prod_{i = 1}^d \lambda_{i, K}})^{-2/d}$, which ensures the consistency with the isotropic case.

\subsubsection{From the estimator to the metric}
As a next step, estimator \eqref{globestim}-\eqref{locestim} is exploited to predict the new spacing of the mesh, known as metric~\cite{freygeorge2008}. In more detail, we combine two distinct criteria, that is the minimization of the mesh cardinality, $\# \mathcal{T}_h$, for a user-defined accuracy, $\tt TOL$, on the discretization error, and an equidistribution of the error throughout the mesh elements, i.e., $\eta_K^2 = {\tt TOL}^2/ \# \mathcal{T}_h$. With this aim, it is instrumental to scale the local estimator \eqref{locestim} with respect to the patch area $|\Delta_K|$, so that
\begin{equation}\label{equiD}
\eta_K^2 = |\Delta_K| \underbrace{\sum_{i=1}^{d}s_{i,K}\, \Big(\ri^T\, \widehat G_{\Delta_K}(E_\nabla)\,\ri\Big)}_{\F(\{s_{i,K}, \ri\}_{i=1,\ldots,d})} = \displaystyle\frac{{\tt TOL}^2}{\# \mathcal{T}_h} = {\tt constant},
\end{equation}
with $\widehat G_{\Delta_K}(\cdot)$ the scaled matrix $G_{\Delta_K}(\cdot)/|\Delta_K|$, and $ |\Delta_K| = ({\prod_{i = 1}^d \lambda_{i, K}}) \, |\widehat \Delta_K|$, where $\widehat \Delta_K=T_K^{-1}(\Delta_K)$ is the pull-back of the patch $\Delta_K$ via map $T_K$.
\\
As a consequence, in order to minimize the mesh cardinality, we are led to maximize the area of each element, which turns out to be equivalent to minimize the quantity $\F(\{s_{i,K}, \ri\}_{i=1,\ldots,d})$ in \eqref{equiD}. This leads us to solve the following constrained minimization problem for each $K \in \mathcal{T}_h$:
\begin{equation}\label{min_mesh}
\min_{s_{i,K}, \ri} \F(\{s_{i,K}, \ri\}_{i=1,\ldots,d}) \ :
\left\{
\begin{array}{l}
{\bf r}_{i, K} \cdot {\bf r}_{j, K}= \delta_{ij}\\[1mm]
s_{1, K}\ge \ldots \ge s_{d, K}\\[1mm]
s_{1, K} \cdot \ldots \cdot s_{d, K}=1,
\end{array}
\right.
\end{equation}
where $\delta_{ij}$ is the Kronecker symbol and $i, j=1, \ldots, d$. The 
solution to such a problem can be computed in a closed form~\cite{Per,farrell2011anisotropic}, as stated in the result below.
\begin{propo}\label{proposition}
	Let $\{g_i, \gi\}_{i=1,\ldots,d}$ be the eigenpairs associated with
	$\widehat G_{\Delta_K}(E_\nabla)$, with $g_1 \geq \ldots \geq g_d > 0$ 
	and $\left\{\gi\right\}_{i=1,\ldots,d}$ orthonormal vectors.
	Then, the solution to the minimization problem \eqref{min_mesh} is
	\begin{equation}\label{r_O}
	s_{i, K}^* = \bigg({\displaystyle \prod_{i = 1}^d g_{i, K}}\bigg)^{1/d} g_{d + 1 - i}^{-1},
    \quad
	\ri^* = {\bf g}_{d + 1 - i} \quad \mbox{with } i = 1, \ldots, d.
	\end{equation}
\\
By the equidistribution criterion, the optimal values for lengths $\lambda_{i, K}^*$ are
\begin{equation}\label{lambda_O}
\lambda_{i, K}^* = \bigg(\displaystyle  \frac{{\tt TOL}^2}{d\, \# \T_h |\widehat \Delta_K|}\bigg)^{1/d}
\bigg({\displaystyle \prod_{i = 1}^d g_{i, K}}\bigg)^{(d-2)/2d^2} g_{d + 1 - i}^{-1/2}.
\end{equation}
\end{propo}

Proposition \ref{proposition} provides the information necessary to assign the spacing characterizing the new mesh, namely the metric $\mathcal M$. It is standard to define $\mathcal M$ as a piecewise constant tensor on the mesh $\mathcal{T}_h$, given by $\M |_K = (R_K^*)^T (\Lambda_K^*)^{-2} R_K^*$, with $(R_K^*)^T = [\ra^*, \ldots, \rc^*]$ and $\Lambda_K^* = {\rm diag}(\la^*, \ldots, \lc^*)$.
\\
Actually, many mesh generators require a nodewise metric information, $\mathcal{M}_1$, so that the optimal quantities in \eqref{r_O} and \eqref{lambda_O} are averaged throughout the patch of elements associated with each vertex, ${\bf V}$, of the mesh $\T_h$. For further details, we refer the interested reader to~\cite{micheletti2010recovery,Per}.

\subsubsection{Tuning the elements deformation through graded anisotropy}\label{graded_sec}

It is well-known that anisotropic adapted grids represent an ideal tool to track steep gradients in an accurate and computationally cheap way. 
Vice versa, in general, isotropic adapted meshes require a larger number of elements to reach the same sharpness in steep gradient detection, since they allow to tune the element size only.
This consideration justifies the employment of anisotropic adapted meshes to track the material/void interface in the topology optimization design.
\\
Nevertheless, strongly deformed elements might bias the mechanical finite element analysis of the optimized layout, by underestimating the displacement and the compliance of the structure.
For this reason, it is advisable to resort to isotropic grids in order to guarantee a reliable engineering characterization of the optimized designs.

As a consequence, following~\cite{ferro2020c}, we exploit anisotropic elements just as a design tool to deliver layouts exhibiting very smooth contours, while preserving an isotropic tessellation within the optimized structure. This leads us to use a so-called graded mesh, which alternates highly stretched elements along the material/void contour with isotropic triangles/tetrahedra inside the structure. 
Such a choice ensures to have a twofold benefit, namely an efficient design tool and mechanically reliable optimized configurations.
\\
To this aim, we modify the metric $\mathcal{M}_1$ derived in the previous section by the rule
$$
\widetilde{\M}_1({\bf V}) = \Theta ({\bf V}) \M_1 ({\bf V}) + \theta({\bf V}) \quad {\bf V} \in \T_h,
$$
where $\Theta, \theta: \Omega \rightarrow \mathbb{R}^{d \times d}$ are generic tensor functions so that metric $
\widetilde{\M}_1$ is symmetric positive definite.

The assessment in Sections~\ref{numerical_sec}-\ref{sec:3d} is carried out for the choices
$$
\Theta({\bf V}) = (1 - \chi_{\varphi, h}({\bf V})) I, \quad  \theta({\bf V}) = \dfrac{1}{h_{iso}^2} \chi_{\varphi, h} ({\bf V}) I,
$$
where $h_{iso}$ is a user-defined sizing of the isotropic mesh inside the structure.

\subsection{The LEVITY algorithm}
This section formalizes the novel design method proposed in this paper. The level set topology optimization in Algorithm~\ref{level_set_algo} is enriched by the anisotropic mesh adaptation procedure detailed in Section~\ref{aniso_sec}.
The quantity driving the mesh adaptation coincides with a suitable filtered function of the level set $\varphi_h$. Namely, we set $ w_h = g(\varphi_h)$ in \eqref{recovered_gradient}, with
\begin{equation}
g(\cdot) = \frac{\tanh(\beta\,\cdot)}{\tanh(\beta)},\label{adapt_driver}
\end{equation}
and $\beta \in \mathbb{R}^+$. Function $\tanh$ acts on $\varphi_h$ by sharpening the transition between values $-1$ and $1$ of the level set function, according to the selected parameter $\beta$. This choice is instrumental in order to confine the mesh adaptation to a thin region around the zero-isocountor of function $\varphi_h$, i.e., in a neighbourhood of the layout boundary $\partial \Sigma$.

The idea is to alternate in a sequential way the topology optimization with the adaptation of the computational mesh. In more detail, Algorithm~\ref{level_set_algo} is enhanced by a new routine which implements the generation of the optimal metric $\M_1$ (or $\widetilde{\M}_1$) in Section~\ref{aniso_sec}. The adaptation step actually takes place only when the percentage variation on the compliance is below a certain threshold (namely, the evolution process is close to converge) or after a fixed number of iterations to overcome the possible slow convergence of the compliance. This strategy avoids to massively increase the computational effort since confining mesh adaptation only to certain iterations, provided that the time step $\Delta t$ is chosen sufficiently small in order to reliably track the evolving contour $\partial \Sigma$.

The combination of the level set approach with mesh adaptivity supports the name LEVITY (LEVel set with mesh adaptivITY) assigned to the proposed algorithm.

\begin{algorithm}[H]
	\caption{LEVITY: LEVel set with mesh adaptivITY}\label{lev_a}
	{\bf Input} :  {\tt CTOL, TOL, ATOL, kmax, kStart, kAdapt}, $\varphi_{h}^{\tt 0}$, $\T_h^{\tt 0}$, $\tt grade$, $h_{iso}$, $\Delta t$, $\alpha$, $V_0$, $\chi_{min}$, $\tau$, $\beta$
	\begin{algorithmic}[1]
		\State Set: ${\tt k} = {\tt 0}$, errComp = 1+{\tt CTOL}, errMesh = 1 + {\tt ATOL} \vspace{1mm}
			\While {errMesh $>$ {\tt ATOL} \& ${\tt k} <$ {\tt kmax}} \vspace{2mm}
            \State $\chi_{\varphi,h}^{\tt k}$ = {\tt characteristic}($\varphi_h^{\tt k}$,
            $\T_h^{\tt k}$
            $\chi_{min}$);\label{char_al_a} \vspace{2mm}
			\State ${\bf u}_h^{\tt k+1}$ = {\tt solveState}($\chi_{\varphi,h}^{\tt k}$, ${\bf v}_h$);\label{state_al_a}\vspace{1mm}
			\State [$d_{t}\overline{F}^{\tt k+1}$, $\kappa$] = {\tt topologicalDerivative}(${\bf u}_h^{\tt k+1}$, $\alpha$, $V_0$);\label{topo_al_a}\vspace{1mm}
			\State $\tilde{\varphi}_h^{\tt k+1}$ = {\tt evolveLevelSet}($\varphi_h^{\tt k}$, $d_{t}\overline{F}^{\tt k+1}$, $\kappa$, $\tau$, $\Delta t$);\label{evolve_al_a}\vspace{2mm}
			\State $\varphi_h^{\tt k+1}$ = {\tt threshold}($\tilde{\varphi}_h^{\tt k+1}$);\label{thr_al_a}\vspace{2mm}
			\State errComp = $|l({\bf u}_h^{\tt k+1})-l({\bf u}_h^{\tt k})|/|l({\bf u}_h^{\tt k})|$;\vspace{2mm}
			\If {${\tt k}>{\tt kStart}$ \& (errComp $< {\tt CTOL}$ $\vert$ ${\tt mod}({\tt k},{\tt kAdapt})==0$)}\label{if_adapt_a}\vspace{2mm}
			\State $\T_h^{\tt k+1}$ = {\tt adaptMesh}($\T_h^{\tt k}$, $\varphi_h^{\tt k+1}$, $\beta$, {\tt TOL}, $\tt grade$,  $h_{iso}$);\label{adapt_a}\vspace{2mm}
			\State  [${\bf u}_h^{\tt k+1}$, $\varphi_h^{\tt k+1}$] = {\tt project}(${\bf u}_h^{\tt k+1}$, $\varphi_h^{\tt k+1}$, $\T_h^{\tt k+1}$);\label{project_a}\vspace{2mm}
			\State errMesh = $|\#\T_h^{\tt k + 1} - \#\T_h^{\tt k}| / |\#\T_h^{\tt k}|$;\label{err_a}\vspace{2mm}
			\Else \vspace{2mm}
			\State $\T_h^{\tt k+1} = \T_h^{\tt k}$;\vspace{2mm}
			\EndIf
			\State {\bf end if} \label{endif_adapt_a} \vspace{2mm}
			\State ${\tt k} = {\tt k+1}$; \vspace{2mm}
			\EndWhile
			\State {\bf end while} \vspace{2mm}
			\State $\overline{\Sigma}_h$ = {\tt extract}($\varphi_h^{\tt k}$);\label{extract_al_a} \vspace{2mm}
			\end{algorithmic}
			{\bf Output}: $\overline{\Sigma}_h$, $\T_h^{\tt k}$
\end{algorithm}

Since LEVITY algorithm represents an enhancement of Algorithm~\ref{level_set_algo}, here we focus on the new input parameters and code blocks.
\\
The new input parameters coincide with the two tolerances $\tt TOL$ and $\tt ATOL$, which control the accuracy of the mesh adaptation according to \eqref{lambda_O} and the stagnation of the mesh cardinality, respectively; the integers $\tt kStart$ and $\tt kAdapt$ that are responsible for switching on the mesh adaptation step; the flag $\tt grade$ which allows us to opt for a fully anisotropic discretization (${\tt grade} = 0$) of the layout or for the graded approach (${\tt grade} = 1$) detailed in Section~\ref{graded_sec}, with a mesh inside the structure of size $h_{iso}$ (for ${\tt grade} = 0$, $h_{iso}$ can be set to any real value, as it is not used); the real value $\beta$ that is used to tune the sharpness of function $g$ in \eqref{adapt_driver}.
\\
Concerning the main changes on the code layout, we remark that the {\bf while} loop now involves a check on the stagnation of the mesh cardinality, in contrast to the control on the stagnation of the compliance in Algorithm~\ref{level_set_algo}. Then, the update of the mesh characterizing the adaptive procedure leads to vary the computational grid employed by the routines in lines \ref{char_al_a}-\ref{thr_al_a}.
\\
The major modification to Algorithm~\ref{level_set_algo} is represented by the command block in lines~\ref{if_adapt_a}-\ref{err_a} through the routines $\tt adaptMesh$ and $\tt project$. After a certain minimum number $\tt kStart$ of iterations is performed, the metric-based methodology in Section~\ref{aniso_sec} is introduced every $\tt kAdapt$ time steps or when the relative variation, errComp, on the compliance is below the user-defined threshold, $\tt CTOL$. The displacement and the level set function are successively projected onto the adapted mesh (line~\ref{project_a}).
\\
Finally, LEVITY algorithm returns the last adapted mesh together with the associated layout $\overline{\Sigma}_h$.

\section{Verification on benchmark $2$D case studies}\label{numerical_sec}

We exploit this section to numerically verify the computational performance of LEVITY algorithm on some two dimensional test cases, i.e., two cantilever and a bridge configurations, which are reference models for topology optimization~\cite{sigmund2004,Valdez17}.
In particular, this section is devoted to assess the reliability of LEVITY and to perform a sensitivity analysis of the output with respect to both the level set and the mesh adaptation processes. 

Since the proposed design tool is independent of the selected medium, we refer to a generic material characterized by a Young modulus and a Poisson ratio equal to $E=1000$ and $\nu=0.3$, respectively.

Concerning Algorithm~\ref{lev_a}, the PDE problems implemented in routines {\tt solveState} and {\tt evolveLevelSet} are approximated with FreeFEM~\cite{freefemmanual}; the metric-based anisotropic mesh adaptation in routine {\tt adaptMesh} is carried out through the dedicated FreeFEM function.

\subsection{Central loaded cantilever}
 As a first benchmark test case, we consider a central loaded cantilever (CLC). The design domain $\Omega$ coincides with the Cartesian set $(0, 2) \times (0,1)$. The external traction ${\bf t} = (0, -5)^T$ is applied to the portion $\Gamma_t=\{(x,y): x=2, 0.45\le y \le 0.55\}$ of the boundary $\partial \Omega$, while a homogeneous Dirichlet data is assigned to ${\bf u}$ on $\Gamma_D=\{(x,y): x=0, 0\le y \le 1\} \subset \partial \Omega$; a homogeneous Neumann boundary condition on $\partial \Omega \setminus (\Gamma_t \cup \Gamma_D)$ completes the physical setting.

\begin{figure}[h]
\begin{center}
\includegraphics[width=0.45\columnwidth]{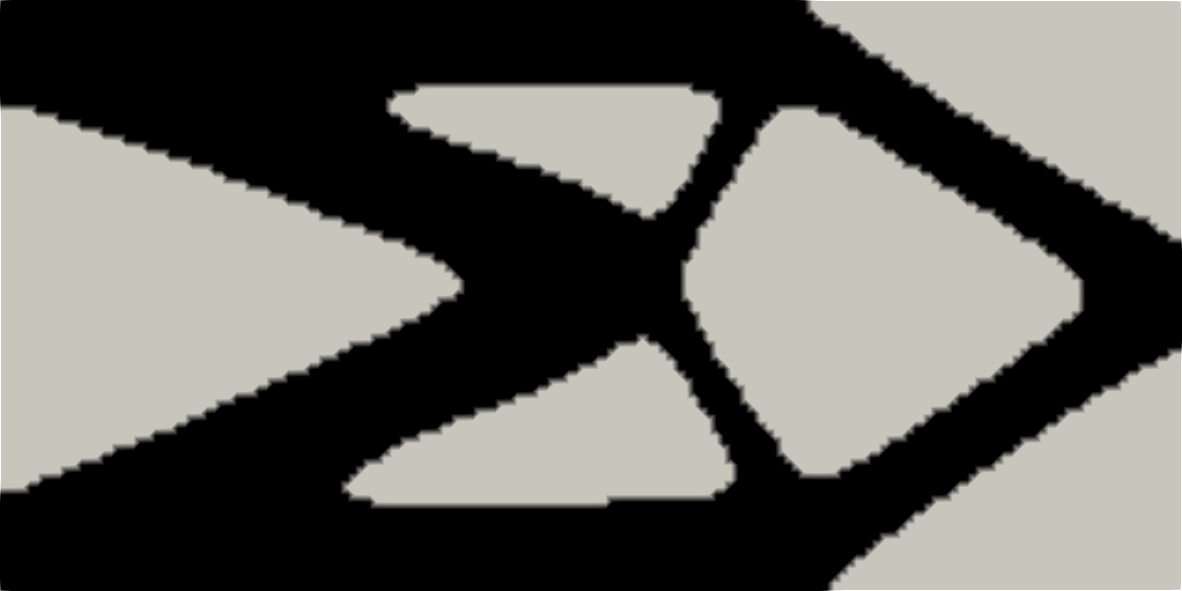}
\includegraphics[width=0.45\columnwidth]{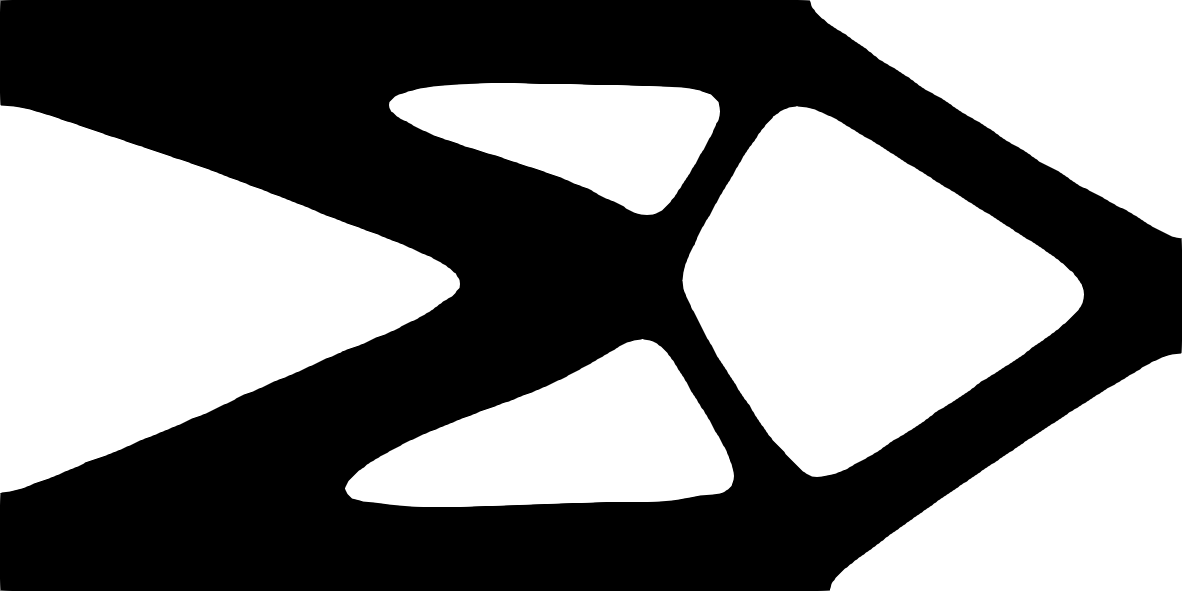}
\caption{CLC. Output of the level set method on a fixed mesh: function $\chi_{\varphi, h}$ (left) and final layout $\overline{\Sigma}_h$ (right).}\label{fig:fixed_clc}
\end{center}
\end{figure}

On this configuration, we solve the minimum compliance problem \eqref{level_set} by setting $\alpha=0.5$, $V_0 = 2$, and $\chi_{min}=1$e-$03$. The level set process is characterized by the choices ${\tt CTOL} = 1$e-$04$; ${\tt kmax}=400$;
the initial guess 
for the level set $\varphi_h^0=1 - 2 \mathbbm{1}_{\Upsilon 1} - 2 \mathbbm{1}_{\Upsilon 2}$,
with 
$$
\begin{array}{rcl}
\Upsilon_1=\{(x,y):  (x-0.5)^2+(y-0.5)^2 \le 0.25^2 \}\\[3mm]
\Upsilon_2=\{(x,y): (x-1.5)^2+(y-0.5)^2 \le0.25^2  \}
\end{array}
$$ 
and $\mathbbm{1}$ the indicator function;
an initial uniform mesh 
$\T_h^{\tt 0}$ consisting of $25600$ triangles; $\Delta t = 0.1$;
$\tau=6$e-$04$.
\\
Algorithm~\ref{level_set_algo} converges in $154$ iterations and returns the characteristic function $\chi_{\varphi,h}$ and the final layout $\overline{\Sigma}_h$ shown in Figure~\ref{fig:fixed_clc}. It is evident that the employment of a fixed computational mesh, which does not match the layout boundary, results in a very irregular trend for function $\chi_{\varphi,h}$. Despite that, the contour of the final CLC inherits the smoothness of the level set function, thus leading to a regular final design characterized by a compliance $l({\bf u}_h)$ equal to $1.40$e-$02$. This value is assumed as the ground truth for the analysis below in terms of mechanical performance.

\begin{figure}[h]
\begin{center}
\includegraphics[width=0.45\columnwidth]{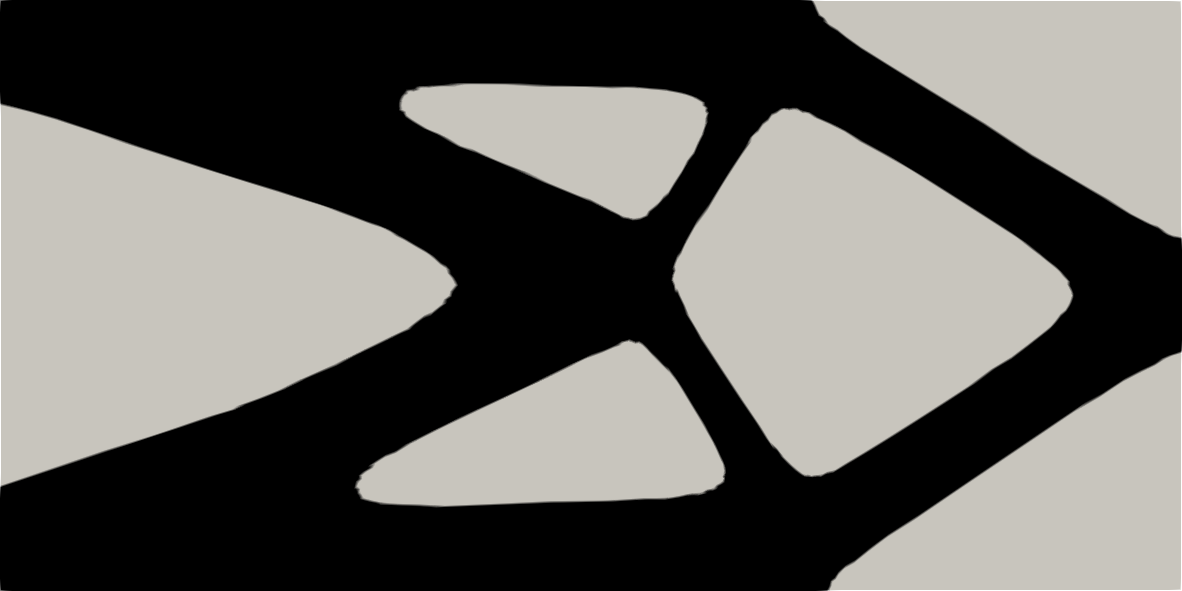}
\includegraphics[width=0.45\columnwidth]{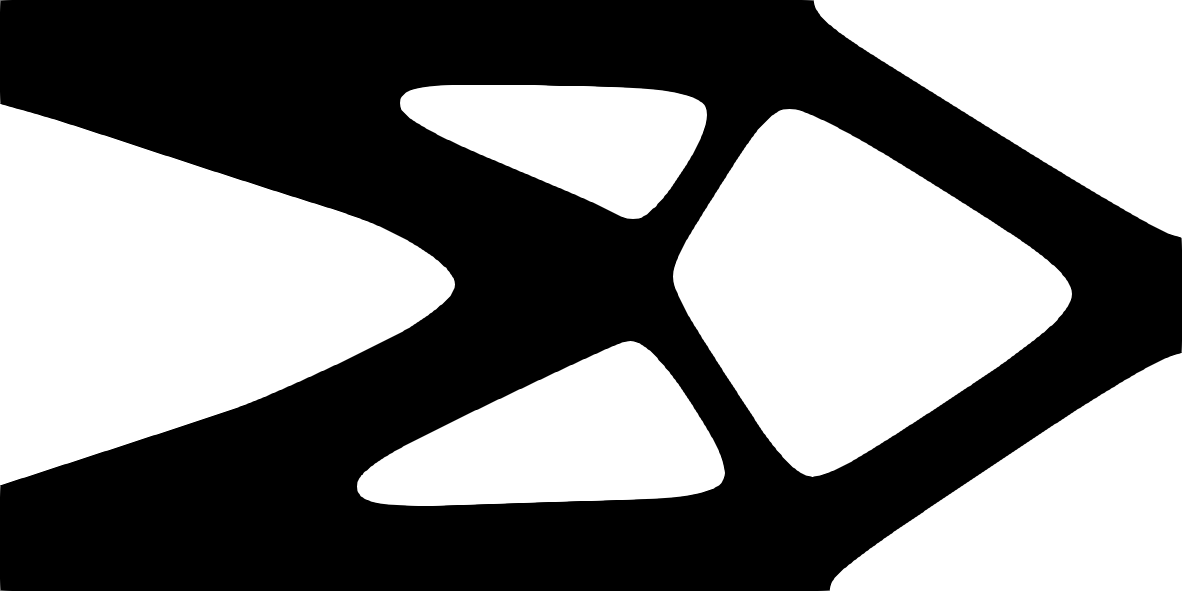}
\caption{CLC. Output of LEVITY algorithm: function $\chi_{\varphi, h}$ (left) and final layout $\overline{\Sigma}_h$ (right).}\label{fig:aniso_clc}
\end{center}
\end{figure}

\begin{figure}[h]
\begin{center}
\includegraphics[width=0.45\columnwidth]{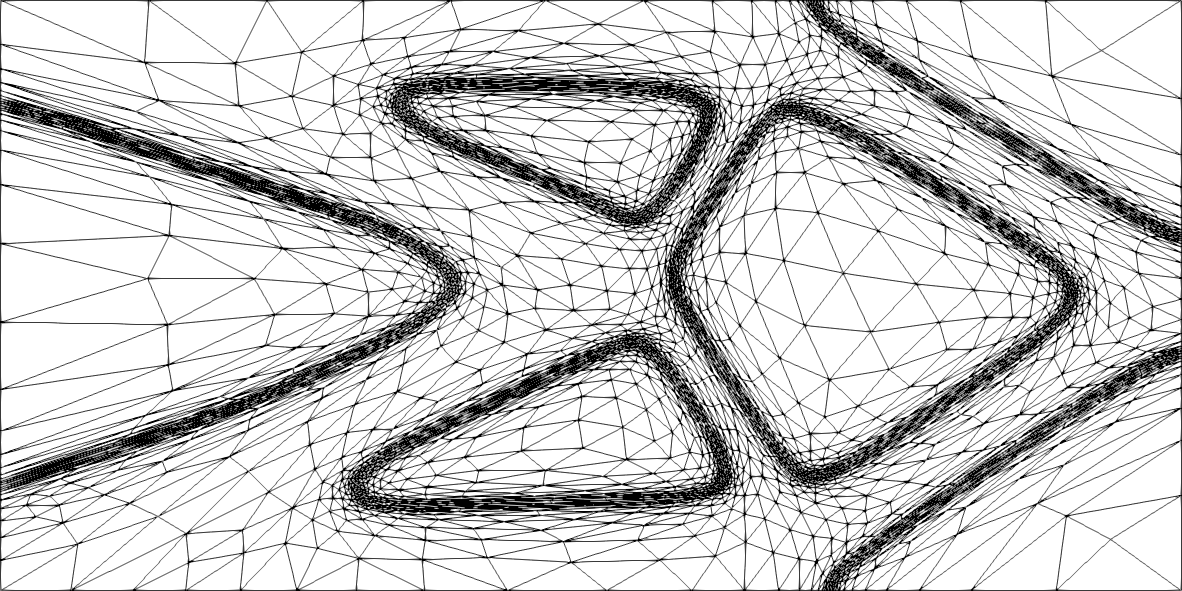}
\includegraphics[width=0.45\columnwidth]{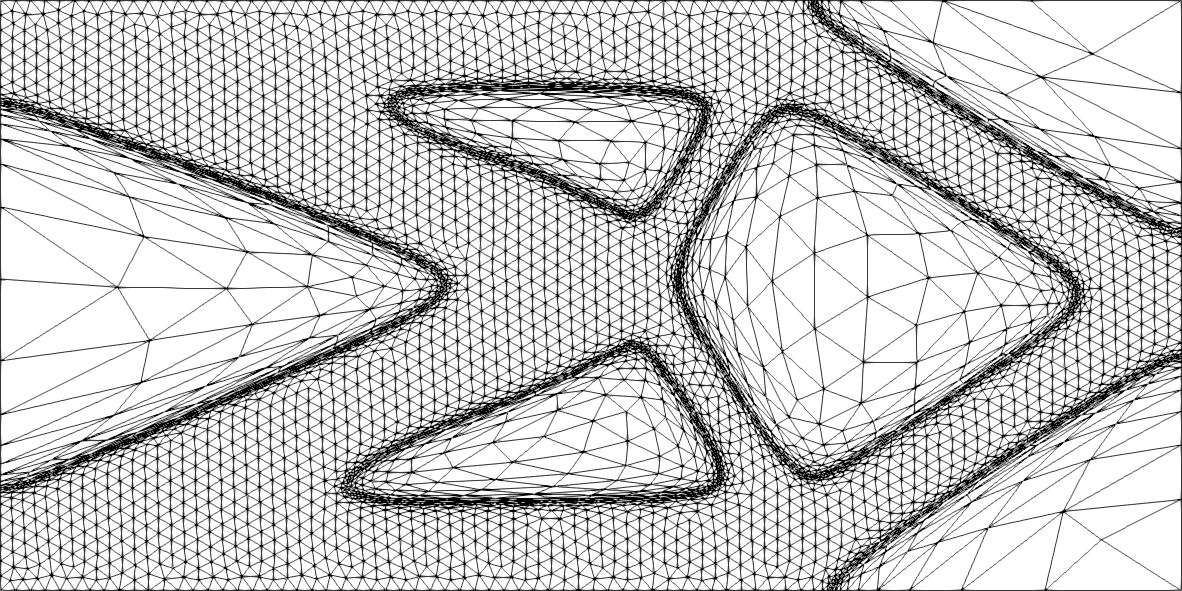}
\caption{CLC. Output tessellation provided by LEVITY algorithm: fully anisotropic (left) and graded (right) mesh.}\label{fig:mesh_clc}
\end{center}
\end{figure}

The improvement led by the mesh adaptation onto the smoothness of the material/void interface identified by $\chi_{\varphi, h}$ can be appreciated in Figure~\ref{fig:aniso_clc}. The two panels show the characteristic function (left) and the structure $\overline{\Sigma}_h$ (right) designed by Algorithm~\ref{lev_a} for the input parameters ${\tt TOL}=8$e-$02$, ${\tt ATOL}=5$e-$03$, ${\tt kStart}=150$, ${\tt kAdapt}=15$, ${\tt grade}=0$, $h_{iso} = 8.0$, and $\beta=10$.
Algorithm~\ref{lev_a} converges after $180$ iterations, while only $5$ mesh adaptation steps take place, with the generation of the adapted tessellation in the left panel of Figure~\ref{fig:mesh_clc}. This grid consists of $5593$ triangles, characterized by a maximum aspect ratio $\max_{K} s_{1, K}$ equal to $82.79$. The most elongated elements are confined to the layout boundary in order to sharply detect the material/void interface. The internal part of the structure is tessellated by anisotropic elements as well, the flag ${\tt grade}$ being set to zero. This feature is likely responsible for the underestimation of the structural compliance $l({\bf u}_h)$ which is $1.19$e-$02$. The poorly reliable mechanical performance associated with a fully anisotropic grid is evident when we switch the grading of the mesh on, by setting ${\tt grade} = 1$ and, for instance, $h_{iso}=1/40$. As shown in Figure~\ref{fig:mesh_clc} (right), the adoption of such a variant of Algorithm~\ref{lev_a} yields a uniform discretization of the internal portion of the structure, which leads to an improvement on the computation of the compliance, now being $l({\bf u}_h) = 1.39$e-$02$. In particular, the graded mesh consists of $7877$ elements with a maximum value for $s_{1, K}$ equal to $34.89$.

Table~\ref{tab_mindisp} provides a further mechanical comparison between the fixed mesh and the two adapted grids in terms of the minimum value of the $y$-component of the displacement, $u_{2, h}^{min}$, (notice that the  maximum quantity, $u_{2, h}^{max}$, is zero, due to the homogeneous boundary data and to the downward orientation of the load).
The layouts associated with the fixed and with the graded meshes are characterized by a very similar value for such a component, while the fully anisotropic case leads to an underestimation of the displacement along the $y$-direction. This quantitative analysis supports that graded grids represent a more mechanically reliable choice with respect to a fully anisotropic discretization of the computational domain. 
\begin{table}[H]
\centering
\begin{tabular}{c | c | c | c} 
 \hline
   & Fixed & Anisotropic & Graded \\
 \hline
 $u_{2, h}^{min}$ & $-2.80$e-$2$ & $-2.49$e-$2$ & $-2.79$e-$2$ \\ 
 \hline
\end{tabular}
\caption{CLC. Minimum value of the displacement along the $y$-direction for the layout optimized on different meshes.}\label{tab_mindisp}
\end{table}

\begin{figure}[h]
\begin{center}
\includegraphics[width=0.45\columnwidth]{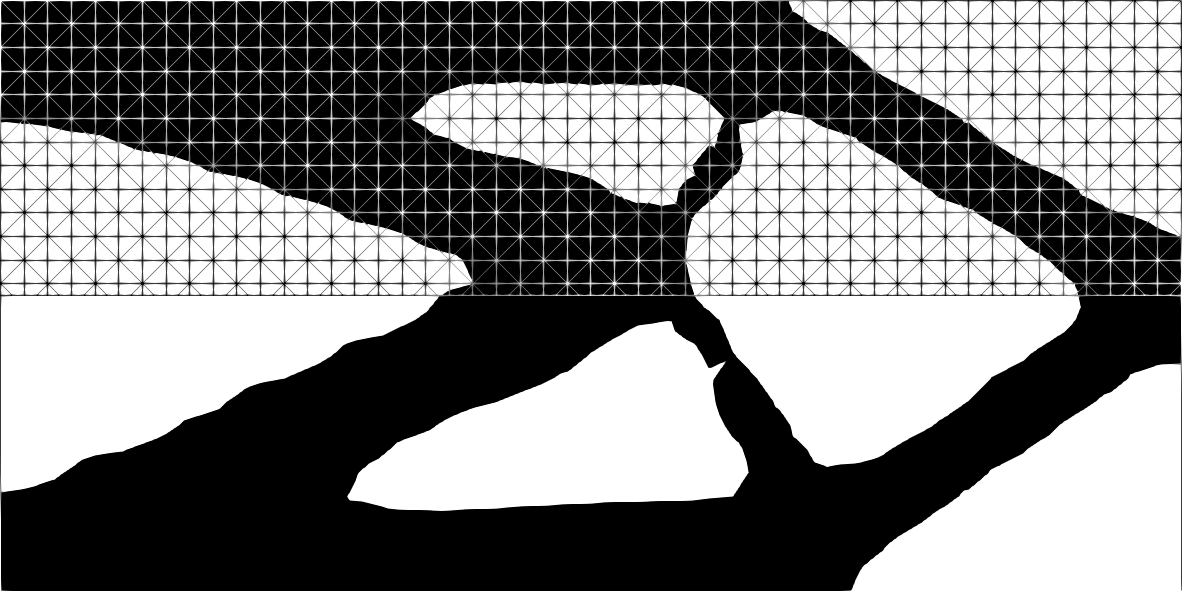}
\includegraphics[width=0.45\columnwidth]{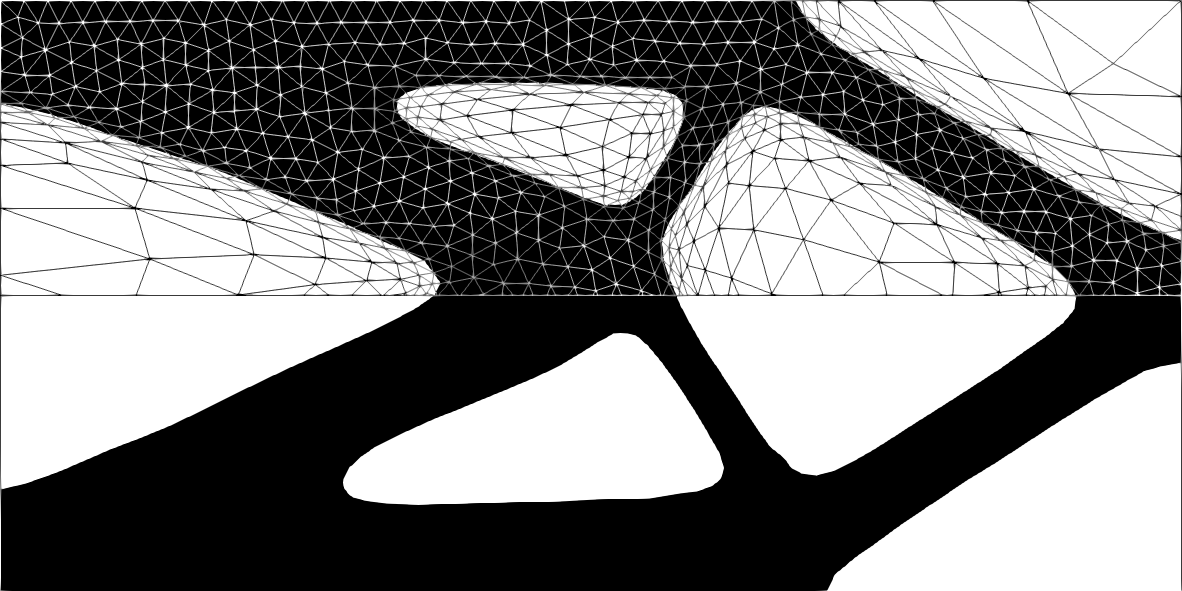}
\caption{CLC. Comparison of the output layout and associated mesh by the level set method (left) and the LEVITY algorithm (right) for $\mathcal{T}_h^{\tt 0} = \mathcal{T}_h^{\tt c}$. }\label{fig:coarse_clc}
\end{center}
\end{figure}

As a further check, we compare the outcome of the level set method when applied to a really coarse grid, by distinguishing between a fixed and a graded domain discretization. 
With this aim, we run Algorithms~\ref{level_set_algo} and~\ref{lev_a} starting from the same initial structured mesh, $\mathcal{T}_h^{\tt c}$, consisting of $2500$ elements. The first procedure yields the layout shown in Figure~\ref{fig:coarse_clc} (left). The mesh is excessively coarse to identify an acceptable structure (notice the thin broken diagonal struts). Moreover, the layout boundary is highly irregular and may require a post-processing phase. On the contrary, the adoption of the graded mesh module allows us to allocate the mesh elements in a strategic way at the expense of a slight increment of the mesh cardinality, now equal to $2654$. As a consequence, the optimized structure is now admissible and exhibits a sharply detected material/void interface.
Despite the employment of a very coarse mesh, the mechanical performance of the optimized CLC is acceptable, with a relative error on the compliance equal to $2.20 \%$ with respect to the ground truth value ($1.37$e-$02$ versus $1.40$e-$02$).

All these considerations lead us to select LEVITY algorithm with graded meshes as the tool for the design of innovative and mechanically reliable topologically optimized structure in the assessment below.

\subsection{Central-bottom loaded bridge}
The second case study coincides with the central-bottom loaded bridge (CBLB), optimized starting from the design domain $\Omega=(-100, 100) \times (0,120)$ when a traction ${\bf t} = (0, -5)^T$ is exerted onto  the boundary portion $\Gamma_t=\{(x,y): -10\le x \le 10, y=0 \}$ of $\partial \Omega$. As additional conditions completing the state equation, we impose $u_2=0$
on $\Gamma_D = \Gamma_{D1} \cup \Gamma_{D2}$, with $\Gamma_{D1}=\{(x,y): -100\le x \le -90, y=0 \}$ and $\Gamma_{D2}=\{(x,y): 90\le x \le 100, y=0 \}$, and homogeneous Neumann data on the remaining part of the boundary $\partial \Omega \setminus ({\Gamma}_t \cup {\Gamma}_{D})$.

\begin{figure}[h]
\begin{center}
\includegraphics[width=0.45\columnwidth]{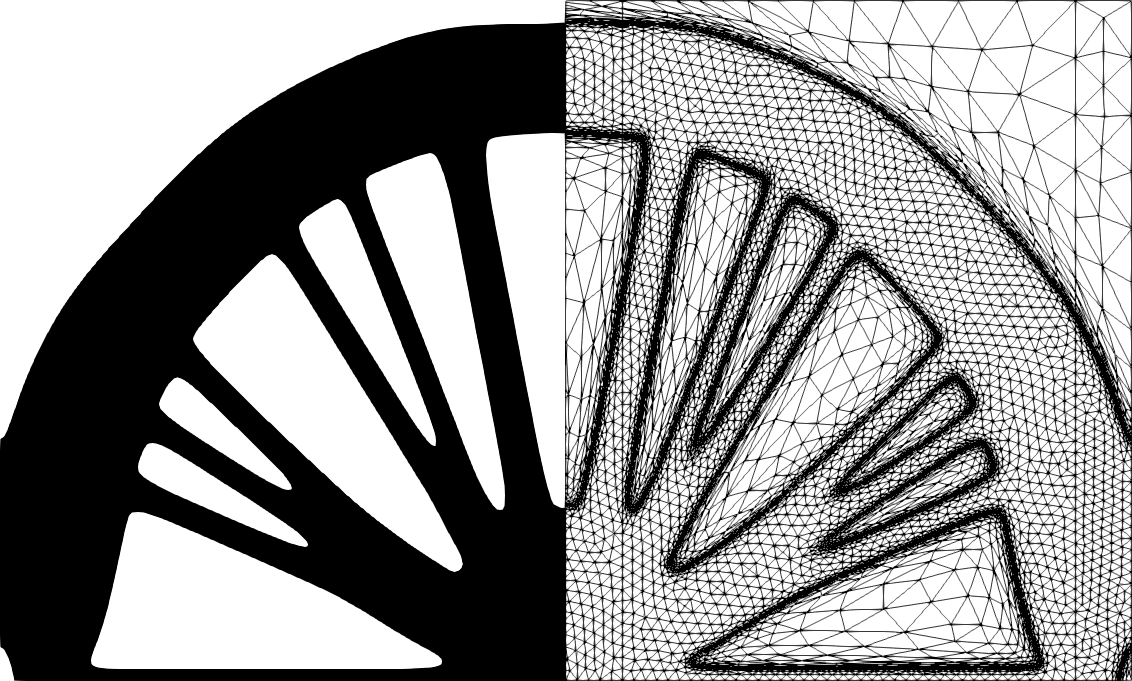}
\includegraphics[width=0.45\columnwidth]{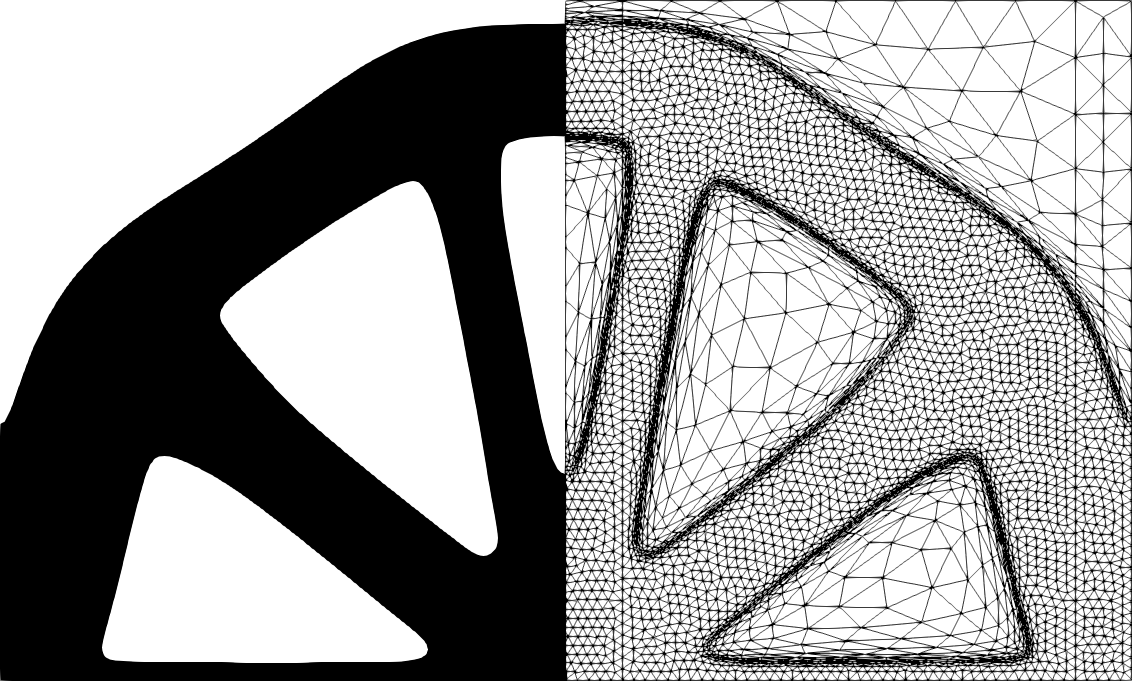}
\caption{CBLB. 
Output layout and associated mesh returned by  LEVITY algorithm for $\tau = 5$e-$01$ (left) and $\tau = 2$ (right). }\label{fig:tau_cblb}
\end{center}
\end{figure}

On this configuration, we run LEVITY with the following choice of the input parameters: ${\tt CTOL} = 1$e-$04$, ${\tt TOL}=3.5$e-$01$, ${\tt ATOL}=5$e-$03$, ${\tt kmax}=400$, ${\tt kStart}=175$, ${\tt kAdapt}=15$, $\varphi_h^0=1$, $\T_h^{\tt 0}$ a uniform mesh with $69120$ elements, ${\tt grade}=1$, $h_{iso}=2$, $\Delta t=0.1$, $\alpha=0.5$, $V_0 = 24000$, $\chi_{min}=1$e-$03$, $\tau=5$e-$01$, and $\beta=10$.
The algorithm requires $217$ iterations to converge, after $6$ mesh adaptations, and delivers the final layout and the mesh shown in Figure~\ref{fig:tau_cblb} (left). The optimized topology associated with this choice of data is rather complex and characterized by a compliance equal to $62.61$. The corresponding graded mesh is made by $22008$ triangles, with very stretched elements confined to the external boundary of the bridge, as well as along the thin struts (with $\max_K s_{1, K} = 72.01$).

\begin{figure}[h]
\hspace*{-.5cm}
\begin{center}
\includegraphics[width=0.48\columnwidth]{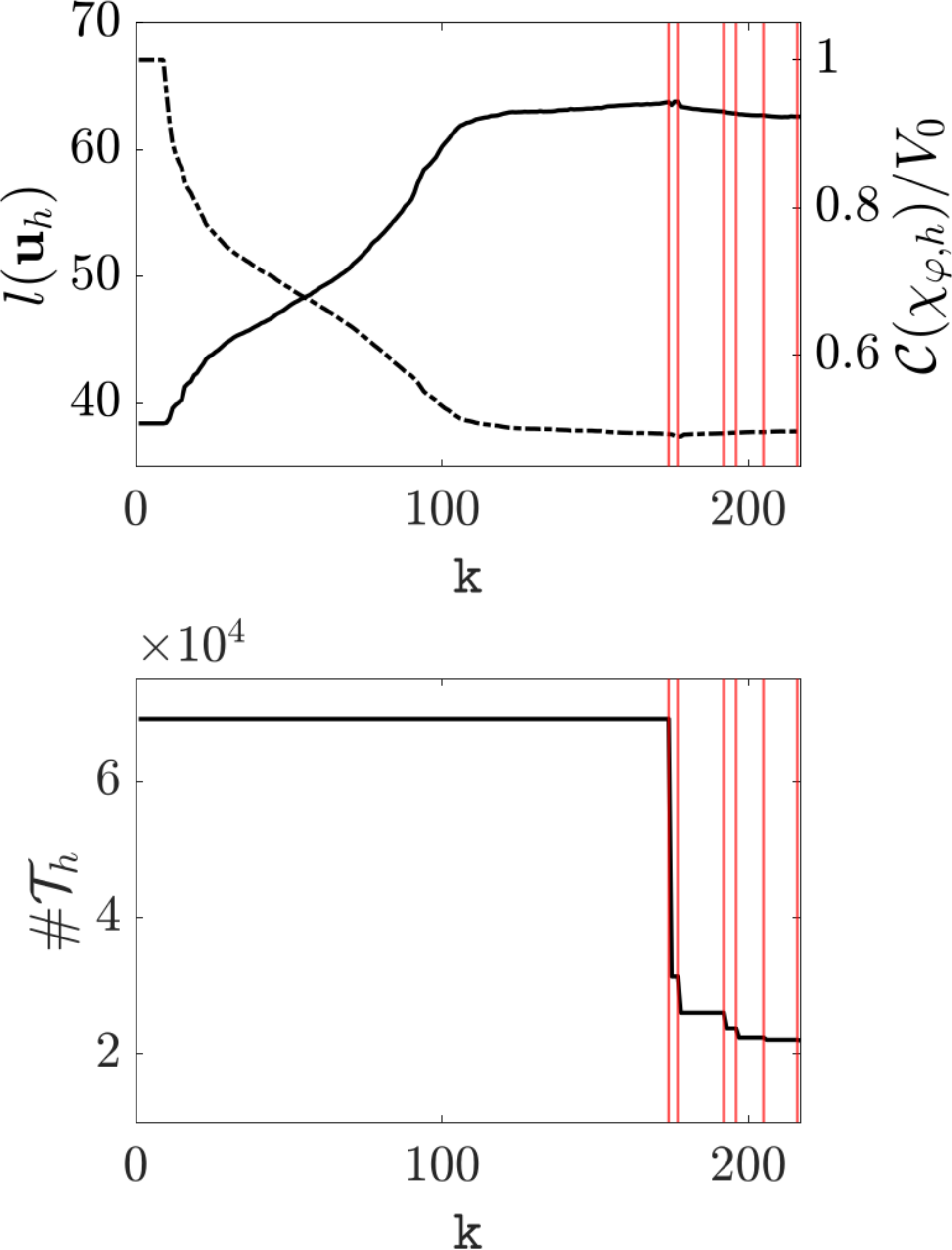}\quad
\includegraphics[width=0.48\columnwidth]{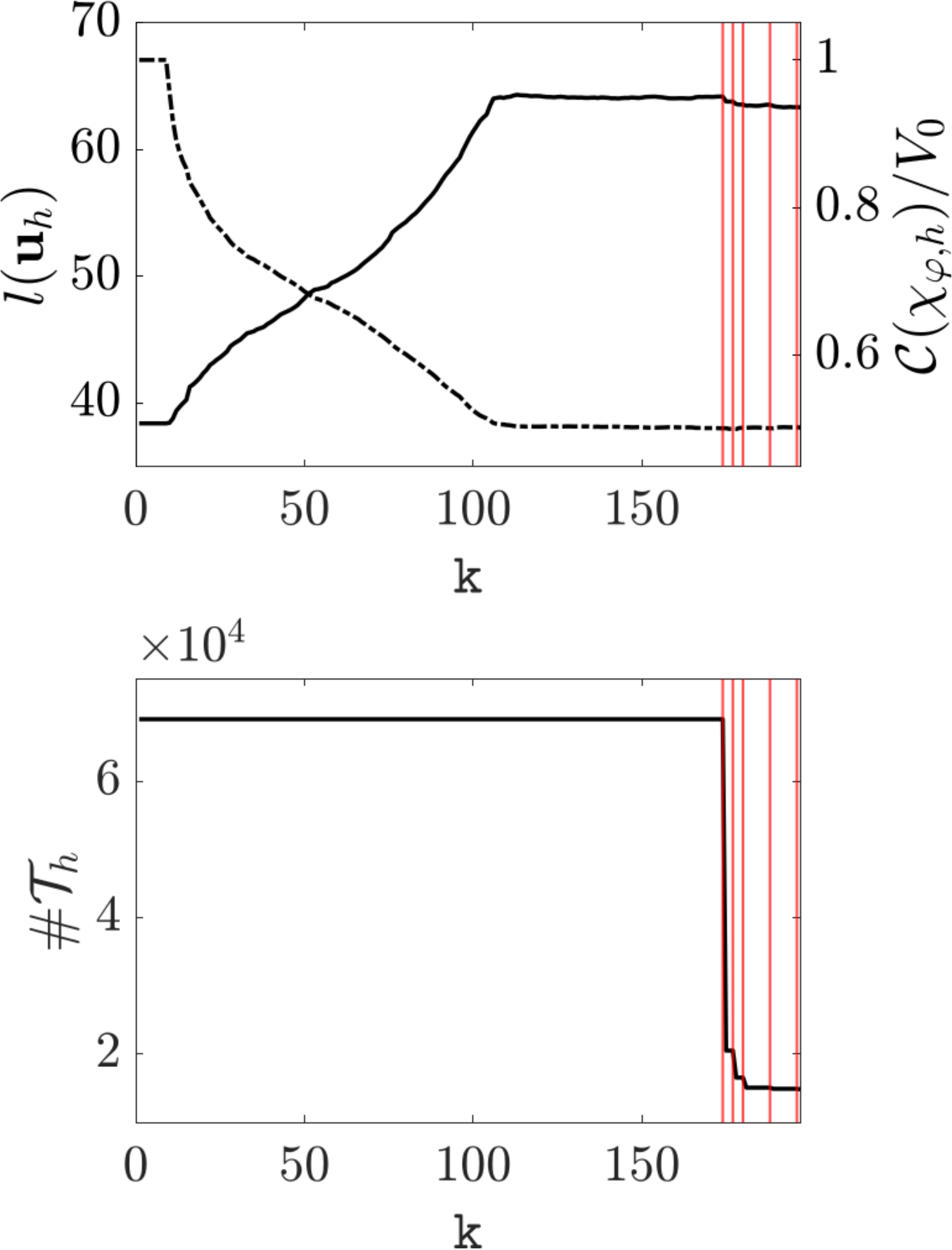}
\caption{CBLB. Convergence history of LEVITY 
for $\tau = 5$e-$01$ (left) and $\tau = 2$ (right): trend of the compliance (solid line) and of the volume fraction (dashed line) (top); evolution of the mesh cardinality (bottom). The vertical lines mark the iterations where mesh adaptation takes place.  }\label{fig:convergence_cblb}
\end{center}
\end{figure}

We exploit this test case to investigate the sensitivity of the output provided by LEVITY to the parameter $\tau$. With this aim, we repeat the previous run by preserving all the input parameters, except for $\tau$ which is now set to $2$. The new structure returned by the algorithm after $197$ iterations, is provided in Figure~\ref{fig:tau_cblb} (right). According to \eqref{evolution}, a larger value of $\tau$ identifies a more diffusive process of the level set evolution. This leads to the design of a structure characterized by a simpler topology with respect to the case $\tau = 5$e-$01$, and by a slightly higher compliance, being $l({\bf u}_h)=63.35$.
\\
The simplified layout is responsible for a coarser graded mesh, with $\# \mathcal{T}_h^{\tt 197} = 14870$, and for a milder maximum element deformation, with $\max_K s_{1, K} = 32.34$. Figure~\ref{fig:convergence_cblb} offers a complete overview of the convergence trend of LEVITY throughout the algorithm iterations in terms of volume fraction and compliance (top) and mesh cardinality (bottom), for both the diffusivities $\tau$. A cross-comparison between the corresponding panels confirms that a small diffusivity in the evolution of the level set leads to a slower convergence history. In particular, the mesh cardinality for $\tau = 5$e-$01$ demands an additional adaptation step to match the stopping criterion, with respect to the case $\tau = 2$. The compliance exhibits a similar trend for both the diffusivities, although the stagnation to the optimized value looks more straightforward for $\tau = 2$.
Finally, the volume fraction decreases in a similar way for both the choices of $\tau$.

As expected, we can conclude that a larger value for $\tau$ leads to the design of structures characterized by a simplified topology and to a faster convergence. 

\subsection{Central loaded short cantilever}

As a last $2$D benchmark case, we solve the optimization problem \eqref{level_set} in the design domain $\Omega=(0, 160) \times (0,128)$, subject to the external traction ${\bf t} = (0, -5)^T$ applied to the boundary portion $\Gamma_t=\{(x,y): x=160, 60\le y\le 68\}$. 
In addition, a null value is assigned to the displacement $\bf u$ on $\Gamma_{D}=\{(x,y): x=0, 0\le y\le 128 \}$ and a homogeneous Neumann boundary condition is imposed on $\partial \Omega \setminus (\Gamma_t \cup \Gamma_{D})$, with $\Gamma_t$, $\Gamma_{D} \subset \partial \Omega$. In the sequel we will refer to this design setting as to the central loaded short cantilever (CLSC).

\begin{figure}[H]
\begin{center}
\includegraphics[width=0.3\columnwidth]{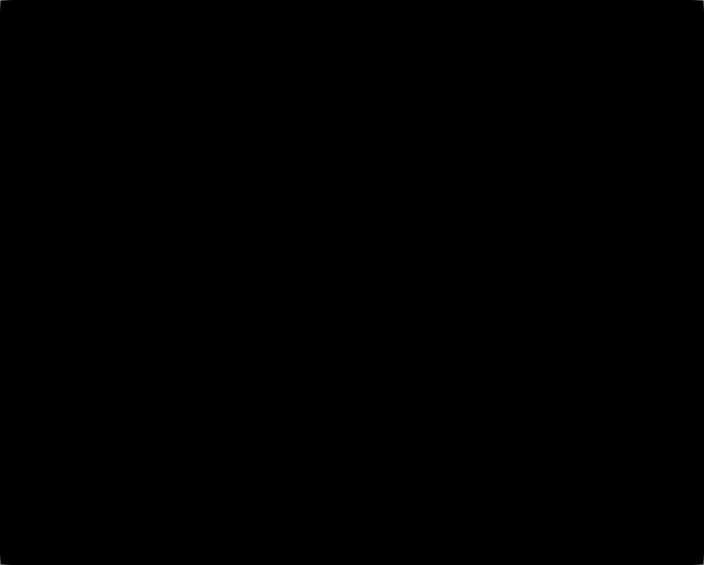}
\includegraphics[width=0.3\columnwidth]{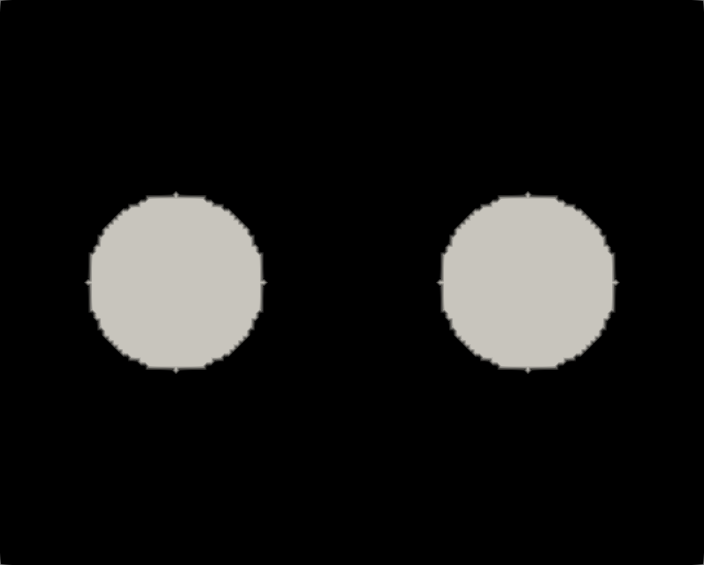}
\includegraphics[width=0.3\columnwidth]{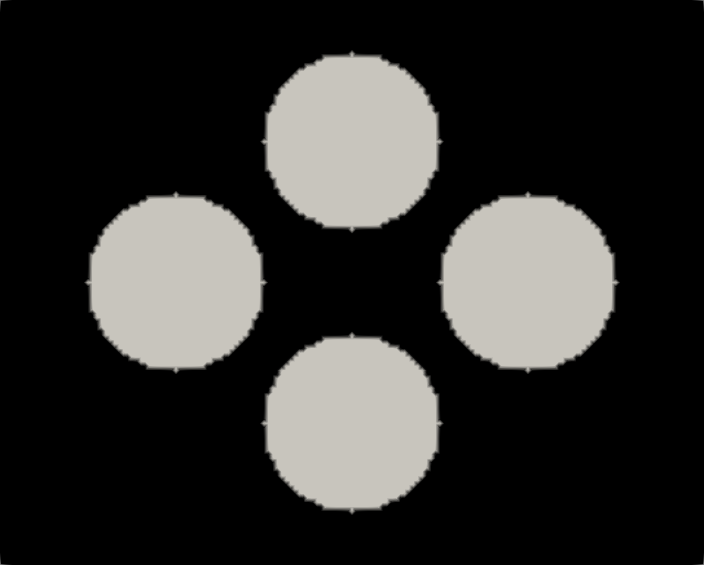}
\\
\includegraphics[width=0.3\columnwidth]{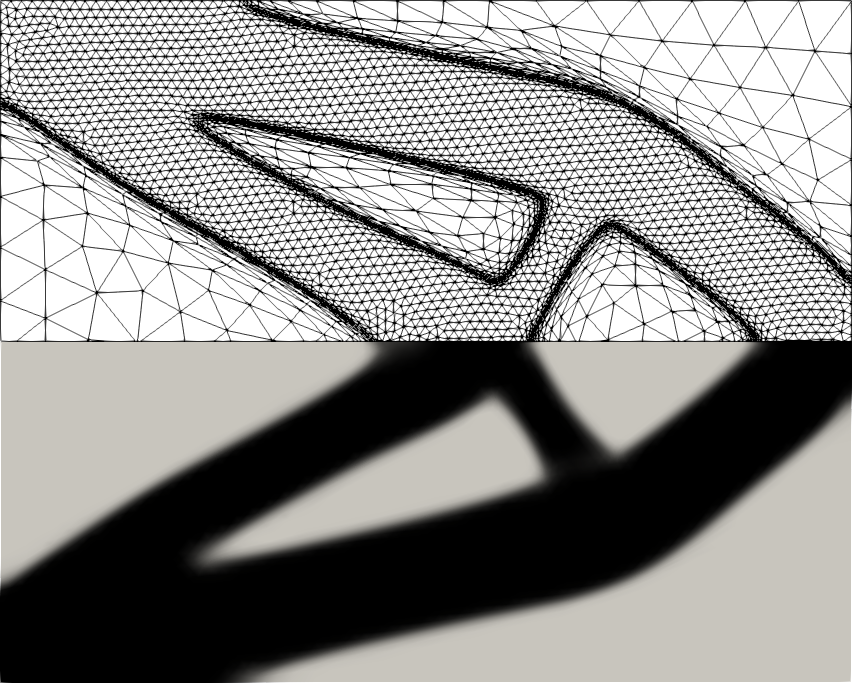}
\includegraphics[width=0.3\columnwidth]{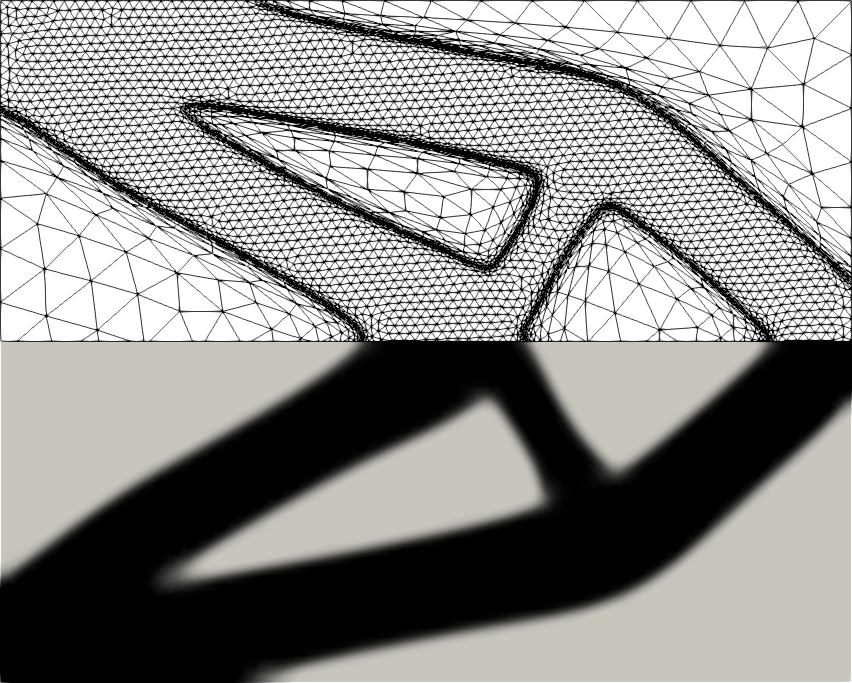}
\includegraphics[width=0.3\columnwidth]{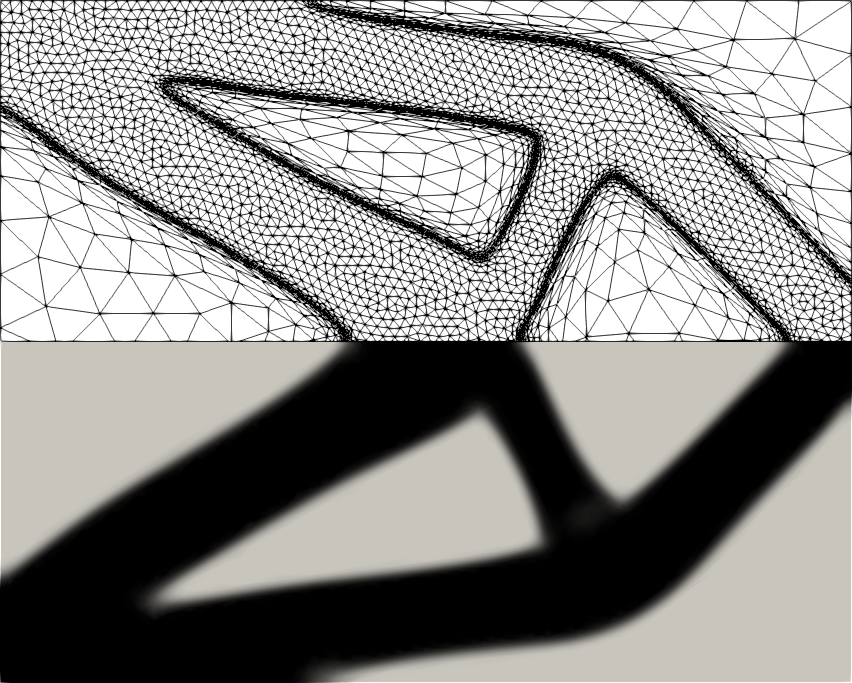}
\caption{CLSC. Sensitivity of the LEVITY algorithm to the initial topology:  $\varphi^0_{1, h}$, $\varphi^0_{2, h}$, $\varphi^0_{3, h}$ (top, left-right) and final level set function overlapped to the associated graded mesh (bottom, left-right).}\label{fig:levels_CLSC}
\end{center}
\end{figure}

Algorithm LEVITY is launched on this configuration, when the input parameters are set to: ${\tt CTOL} = 1$e-$04$, ${\tt TOL}=3.5$e-$01$, ${\tt ATOL}=5$e-$03$, ${\tt kmax}=400$, ${\tt kStart}=175$, ${\tt kAdapt}=15$, $\varphi_{h}^0 = \varphi_{1,h}^0=1$, $\T_h^{\tt 0}=\T_h^{\tt 0, A}$ a uniform mesh consisting of $4960$ elements, ${\tt grade}=1$, $h_{iso}=2$, $\Delta t=0.1$, $\alpha=0.5$, $V_0 = 20480$, $\chi_{min}=1$e-$03$, $\tau=1$, $\beta=10$.
After $201$ iterations and $5$ mesh adaptation steps, we break the {\bf while} loop with the structure, CLSC-L-T1A, associated with the optimized level set function in Figure~\ref{fig:levels_CLSC} (bottom-left), here overlapped to the final graded mesh.

This configuration is exploited to investigate the sensitivity of the output by LEVITY to the initial topology. To this aim, we re-run the algorithm by varying $\varphi_{h}^0$, selected as in the top-center ($\varphi_{2,h}^0$) and the top-right ($\varphi_{3,h}^0$) panels of Figure~\ref{fig:levels_CLSC}. The corresponding level set function and graded mesh are provided in the bottom-center and bottom-right panels of Figure~\ref{fig:levels_CLSC}, which are referred to as CLSC-L-T2A and CLSC-L-T3A, from now on. The output topology is the same. We recognize only a slight variation of the hole size when the initial topology becomes more complex. 
\\
Table~\ref{tab_levsens} gathers some quantitative information about the mechanical performance of the structures  and about the mesh features. We observe a small increment of the compliance when moving from CLSC-L-T1A to CLSC-L-T3A, whereas essentially the same cardinality and maximum aspect ratio characterize the final graded mesh. This confirms the robustness of the LEVITY algorithm with respect to the choice of the initial topology.


\begin{table}[H]
\centering
\begin{tabular}{c | c | c | c} 
 \hline
   & $l({\bf u}_h)$ & $\# {\mathcal T}_h^{\tt k}$ & $\max_K s_{1, K}$ \\
 \hline
 CLSC-L-T1A & $30.58$ & $12534$ & $48.35$\\ 
 CLSC-L-T2A & $31.70$ & $12382$ & $42.81$\\
 CLSC-L-T3A & $31.81$ & $12129$ & $48.89$\\
 \hline
\end{tabular}
\caption{CLSC. Sensitivity of LEVITY algorithm to the initial topology: compliance (first column), cardinality (second column) and maximum aspect ratio (third column) associated with the final graded mesh.}\label{tab_levsens}
\end{table}

\begin{figure}[H]
\begin{center}
\includegraphics[width=0.3\columnwidth]{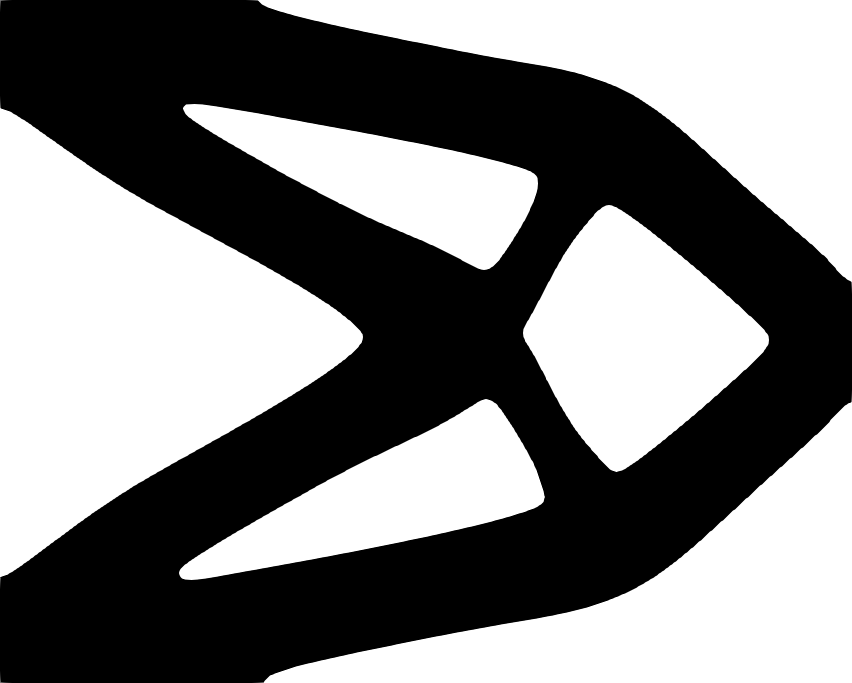}
\includegraphics[width=0.3\columnwidth]{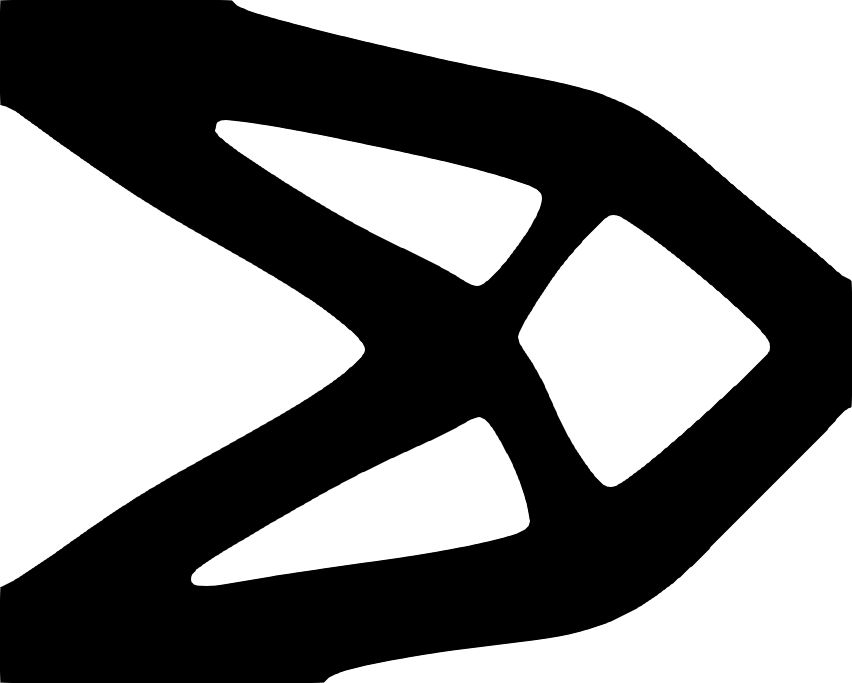}
\includegraphics[width=0.3\columnwidth]{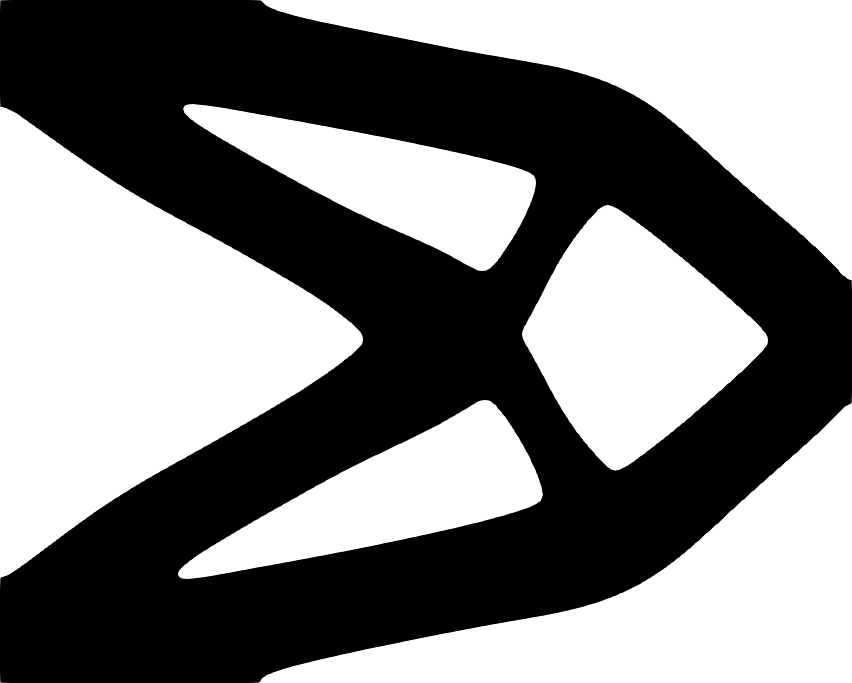}
\caption{CLSC. Sensitivity of the LEVITY algorithm to the initial mesh: final layout $\overline{\Sigma}_h$ for ${\mathcal T}_{h}^{0}$ coinciding with ${\mathcal T}_{h}^{\tt 0, A}$ (left), ${\mathcal T}_{h}^{\tt 0, B}$ (center), and ${\mathcal T}_{h}^{\tt 0, C}$ (right).}\label{fig:sens_mesh_CLC}
\end{center}
\end{figure}

We further explore the sensitivity of LEVITY to the initial mesh by running the algorithm for $\varphi_{h}^0 = \varphi_{2,h}^0$ and ${\mathcal T}_{h}^{\tt 0} = {\mathcal T}_{h}^{\tt 0, A}$, ${\mathcal T}_{h}^{\tt 0, B}$, ${\mathcal T}_{h}^{\tt 0, C}$, with ${\mathcal T}_{h}^{\tt 0, B}$ and ${\mathcal T}_{h}^{\tt 0, C}$ uniform meshes consisting of $2560$ and $92160$ triangles, respectively,  while preserving all the other input parameters. Figure~\ref{fig:sens_mesh_CLC} and Table~\ref{tab_mindisp2} summarize the results of such a comparison. It turns out that LEVITY output is essentially independent of the initial mesh also, since it provides the same final layout, as well as very similar values for the quantities tracked in the table.
\begin{table}[H]
\centering
\begin{tabular}{c | c | c | c} 
 \hline
   & $l({\bf u}_h)$ & $\# {\mathcal T}_h^{\tt k}$ & $\max_K s_{1, K}$ \\
 \hline
 CLSC-L-T2A & $31.70$ & $12382$ & $42.81$\\
 CLSC-L-T2B & $30.77$ & $11838$ & $44.65$\\
 CLSC-L-T2C & $31.22$ & $11733$ & $41.99$\\
 \hline
\end{tabular}
\caption{CLSC. Sensitivity of LEVITY algorithm to the initial mesh: compliance (first column), cardinality (second column) and maximum aspect ratio (third column) associated with the final graded mesh.}\label{tab_mindisp2}
\end{table}

The improvements due to anisotropic adapted meshes in a topology optimization process have been already assessed when dealing with the most widespread density-based method, namely the SIMP (Solid Isotropic Material with Penalization)~\cite{bendsoe1995, sigmund2004, rozvany}. In particular, a new algorithm, named SIMPATY (SIMP with mesh AdaptiviTY), has been proposed in~\cite{soli2019} for fully anisotropic adapted grids and successively modified to account for graded meshes~\cite{ferro2020b,ferro2020c}. The main difference between SIMPATY and LEVITY consists in the procedure adopted to change the topology under optimization. In particular, in order to track the optimized layout boundary, SIMPATY replaces the evolution equation \eqref{evolution} with a sound optimization step. It has been verified that SIMPATY algorithm provides original free-form optimized designs, and significantly limits the standard post-processing required by the SIMP layouts.
Indeed, anisotropic adapted meshes remove any staircase effect with a sharp detection of the layout boundary. In addition, SIMPATY algorithm effectively erases checkerboard effects, despite the adoption of the same space to discretize both the displacement, $\bf u$, and the density function, $\rho$. Further developments of SIMPATY algorithm include the setting of a new process at the microscale for the design of innovative metamaterials in a multi-objective, multi-physics, multi-scale setting~\cite{ferro2020,Cristofaro2021,Ferro2021,Gavazzoni2022}, as well as of a reduced order model for topology optimization~\cite{ferro2019}.

\begin{figure}[H]
\begin{center}
\includegraphics[width=0.3\columnwidth]{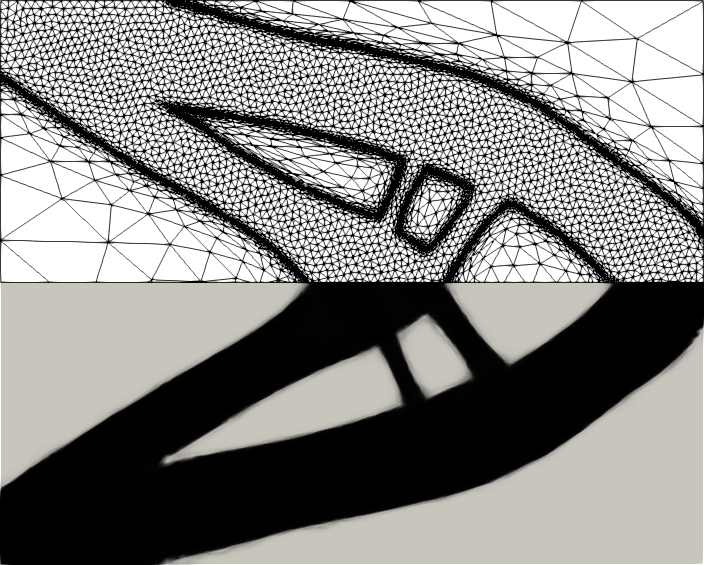}
\includegraphics[width=0.3\columnwidth]{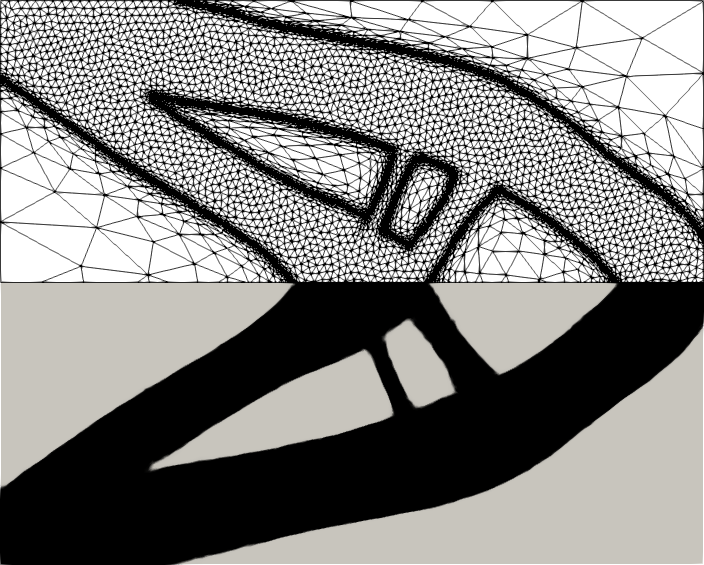}
\includegraphics[width=0.3\columnwidth]{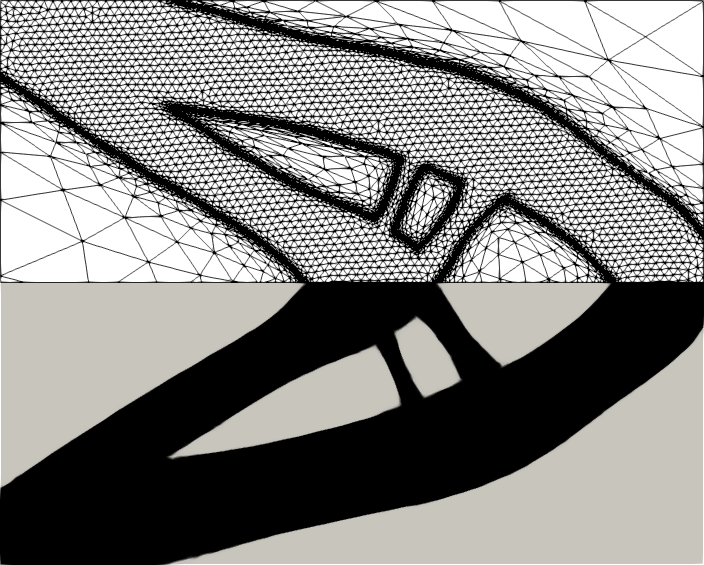}
\caption{CLSC. Sensitivity of the SIMPATY$_G$ algorithm to the initial density: final density function overlapped to the associated graded mesh for $\rho^0_{1, h}$ (left), $\rho^0_{2, h}$ (center), $\rho^0_{3, h}$ (right).}\label{fig:levels_simp}
\end{center}
\end{figure}

Despite the different approaches at the base of SIMP and level set methods, it may be of interest to check the performance of SIMPATY algorithm when applied to the CLSC configuration. With this aim, we also consider a variant of SIMPATY algorithm, named SIMPATY$_G$, where the standard functional $l({\bf u})$ is replaced by
\begin{equation}\label{LG}
l_{G}({\bf u}, \rho) = l({\bf u}) + \dfrac{1}{2} \gamma \int_\Omega |\nabla \rho|^2,
\end{equation}
with $\gamma \in \mathbb{R}^+$, which represents a regularization of the compliance \eqref{compliance} with a control on the structure perimeter.
\\
We replicate the simulation in Figure~\ref{fig:levels_CLSC}, for $l_{G}$ with $\gamma = 2.5$e-$02$, and starting from an initial density $\rho^0_{i, h}$, which mimics the initial topology $\varphi^0_{i, h}$, for $i = 1, \ldots, 3$.  These new runs yield the three layouts in Figure~\ref{fig:levels_simp}. 
\noindent
The optimized topology designed by SIMPATY$_G$ is different when compared with LEVITY output and exhibits additional holes.
Nevertheless, similarly to LEVITY algorithm, SIMPATY$_G$ is scarcely sensitive to the initial density distribution, as highlighted by the three panels in Figure~\ref{fig:levels_simp} and by the values in Table~\ref{tab_simpg}.
\begin{table}[H]
\centering
\begin{tabular}{c | c | c | c} 
 \hline
   & $l({\bf u}_h)$ & $\# {\mathcal T}_h^{\tt k}$ & $\max_K s_{1, K}$ \\
 \hline
 CLSC-S-T1A & $30.79$ & $15044$ & $47.9$\\
 CLSC-S-T2A & $30.38$ & $16117$ & $57.143$\\
 CLSC-S-T3A & $30.78$ & $16427$ & $43.24$\\
 \hline
\end{tabular}
\caption{CLSC. Sensitivity of SIMPATY$_G$ algorithm to the initial density: compliance (first column), cardinality (second column) and maximum aspect ratio (third column) associated with the final graded mesh.}\label{tab_simpg}
\end{table}

\begin{figure}[H]
\begin{center}
\includegraphics[width=0.3\columnwidth]{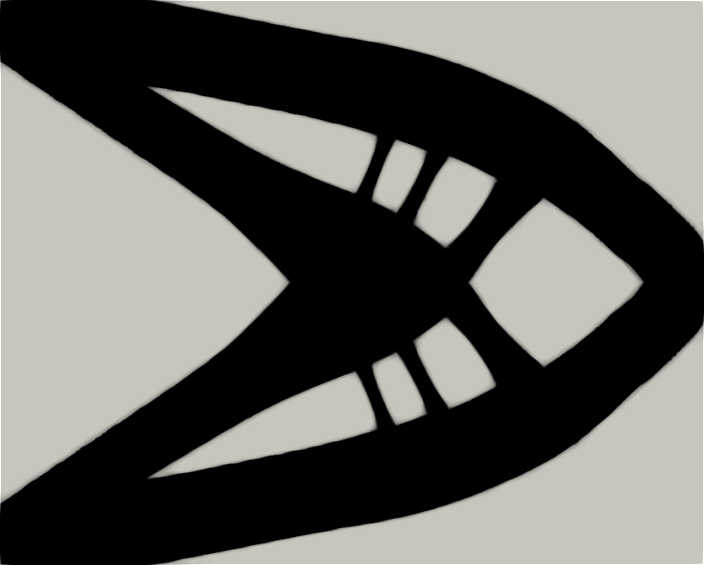}
\includegraphics[width=0.3\columnwidth]{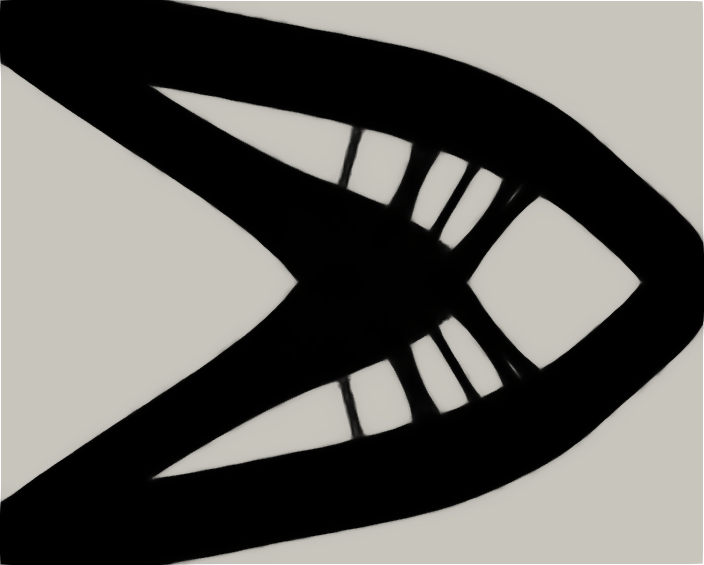}
\includegraphics[width=0.3\columnwidth]{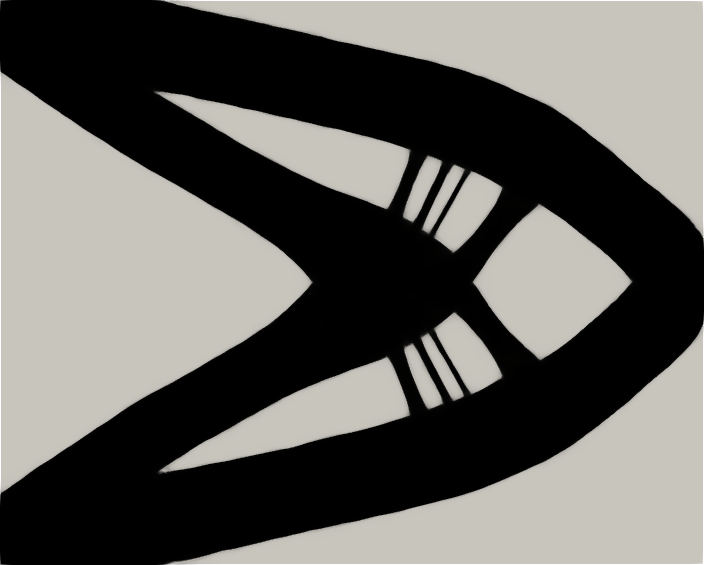}
\caption{CLSC. Sensitivity of the SIMPATY algorithm to the initial density: final density function for $\rho^0_{1, h}$ (left), $\rho^0_{2, h}$ (center), $\rho^0_{3, h}$ (right).}\label{fig:simp_sens}
\end{center}
\end{figure}

We remark that the regularization term in \eqref{LG} is crucial to make SIMPATY independent of the initial topology. This is confirmed by Figure~\ref{fig:simp_sens}, which replicates the same simulations as in Figure~\ref{fig:levels_simp}, by setting $\gamma = 0$,

\section{LEVITY as a design tool for $3$D configurations}\label{sec:3d}

The $3$D tests are employed to check the free-form design capability of LEVITY in two different applications.
In this framework, we still employ FreeFEM to solve the PDE problems involved in Algorithm~\ref{lev_a}, while resorting to Mmg~\cite{mmg} to manage the generation of the graded meshes.

\subsection{The bridge}

The first $3$D topology optimization is applied to the  parallelepiped $\Omega=(0, 6)\times(0, 1)\times(0, 1)$, constituted by a material with unitary Young modulus and null Poisson ratio. The design domain is loaded in the area $\Gamma_t=\{ (x, y, z): 2.95\le x \le 3.05, 0\le y\le 1, z=1\}$ by the traction
${\bf t}=(0, 0, -1)^T$ and subject to homogeneous Dirichlet conditions on the boundary portion $\Gamma_D=\Gamma_{D1}\cup \Gamma_{D2}\cup \Gamma_{D3}\cup \Gamma_{D4}$, with
$$
\begin{array}{rcl}
\Gamma_{D1}&=&\{ (x, y, z) : 0\le x \le 0.01, 0\le y\le 0.01, z=0\},\\[2mm]
\Gamma_{D2}&=&\{ (x, y, z) : 5.99\le x \le 6, 0\le y\le 0.01, z=0\},\\[2mm]
\Gamma_{D3}&=&\{ (x, y, z) : 0\le x \le 0.01, 0.99\le y\le 1, z=0\},\\[2mm]
\Gamma_{D4}&=&\{ (x, y, z) : 5.99\le x \le 6, 0.99\le y\le 1, z=0\},
\end{array}
$$
and to homogeneous Neumann data on $\partial \Omega \setminus (\Gamma_D \cup \Gamma_t)$. This configuration leads to the design of a bridge structure.
\\
For computational reasons, we perform the optimization on a quarter of the domain, and then we recover the whole structure by symmetry.

LEVITY algorithm is run by setting the input parameters as: ${\tt CTOL} = 1$e-$04$, ${\tt TOL}=6.5$, ${\tt ATOL}=5$e-$03$, ${\tt kmax}=500$, ${\tt kStart}=150$, ${\tt kAdapt}=20$, $\varphi_h^0=1$, $\T_h^{\tt 0}$ a uniform mesh with $52556$ tetrahedra, ${\tt grade}=1$, $h_{iso}=0.025$, $\Delta t=0.1$, $\alpha=0.1$, $V_0 = 6$, $\chi_{min}=1$e-$03$, $\tau=4$e-$04$, and $\beta=5$.

\begin{figure}[H]
\begin{center}
\includegraphics[width=0.65\columnwidth]{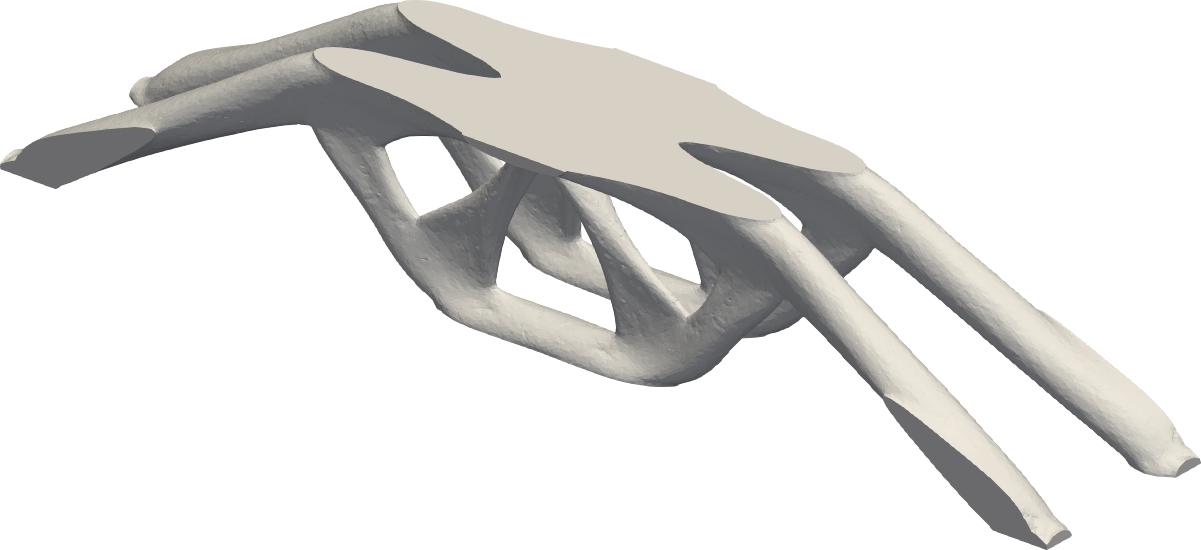}
\\
\includegraphics[width=0.65\columnwidth]{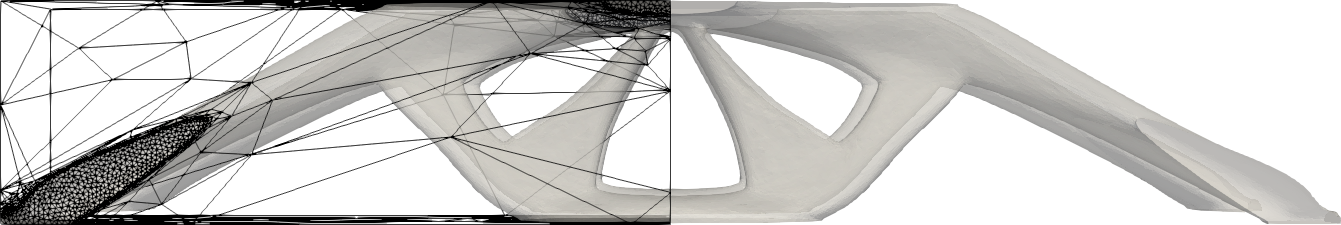}
\caption{The bridge. Output of LEVITY algorithm: final layout $\overline{\Sigma}_h$ (top) and graded mesh superimposed to the optimized structure (bottom).}\label{fig:bridge_levelset}
\end{center}
\end{figure}

\begin{figure}[H]
\begin{center}
\includegraphics[height=0.25\columnwidth]{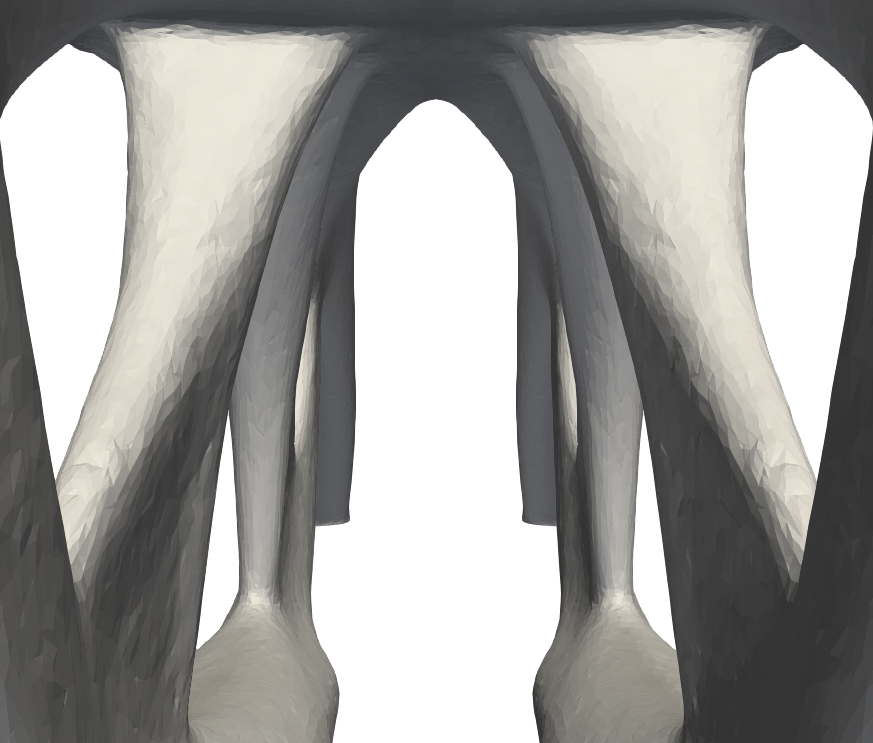}
\includegraphics[height=0.25\columnwidth]{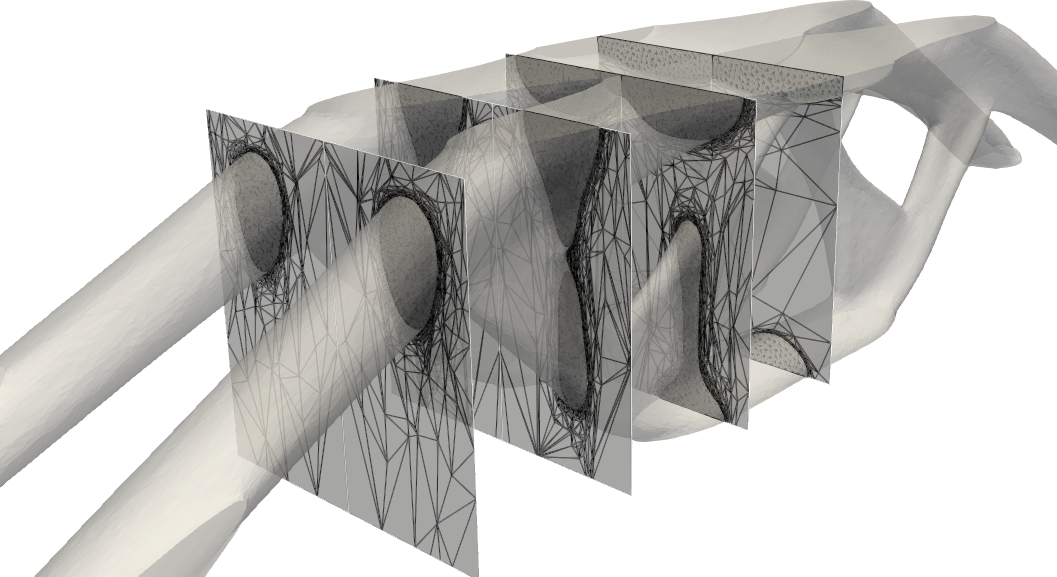}
\caption{The bridge. Details of the output of LEVITY algorithm: spans (left) and slices of the graded mesh (right).}\label{fig:bridge_details}
\end{center}
\end{figure}

Figure~\ref{fig:bridge_levelset} shows the output of Algorithm~\ref{lev_a} provided after $471$ iterations and $17$ mesh adaptations and characterized by a compliance equal to $1.08$. The top panel displays the final layout which turns out to be particularly smooth thanks to the employment of anisotropic tetrahedra along the solid/void structure interface. The bottom panel shows the computational mesh. The graded features of the grid are evident with very regular elements inside the structure (see the bottom-left corner) and a very stretched tessellation outside the bridge. Moreover, we can appreciate the different element density inside and outside the optimized layout.
Further details about the bridge structure and the graded mesh are provided in Figure~\ref{fig:bridge_details}.


\begin{figure}[H]
\begin{center}
\includegraphics[width=0.475\columnwidth]{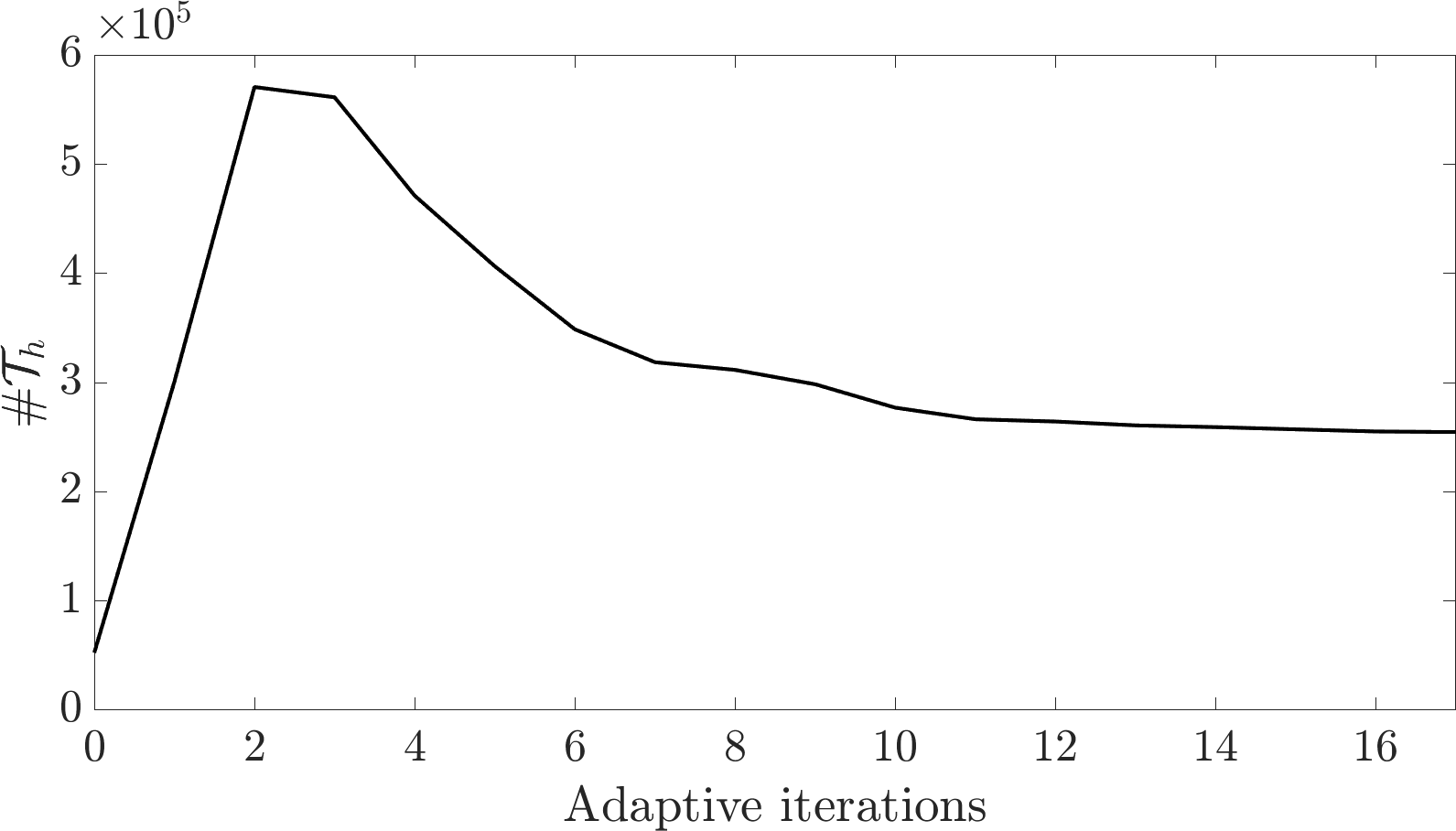}
\quad
\includegraphics[width=0.475\columnwidth]{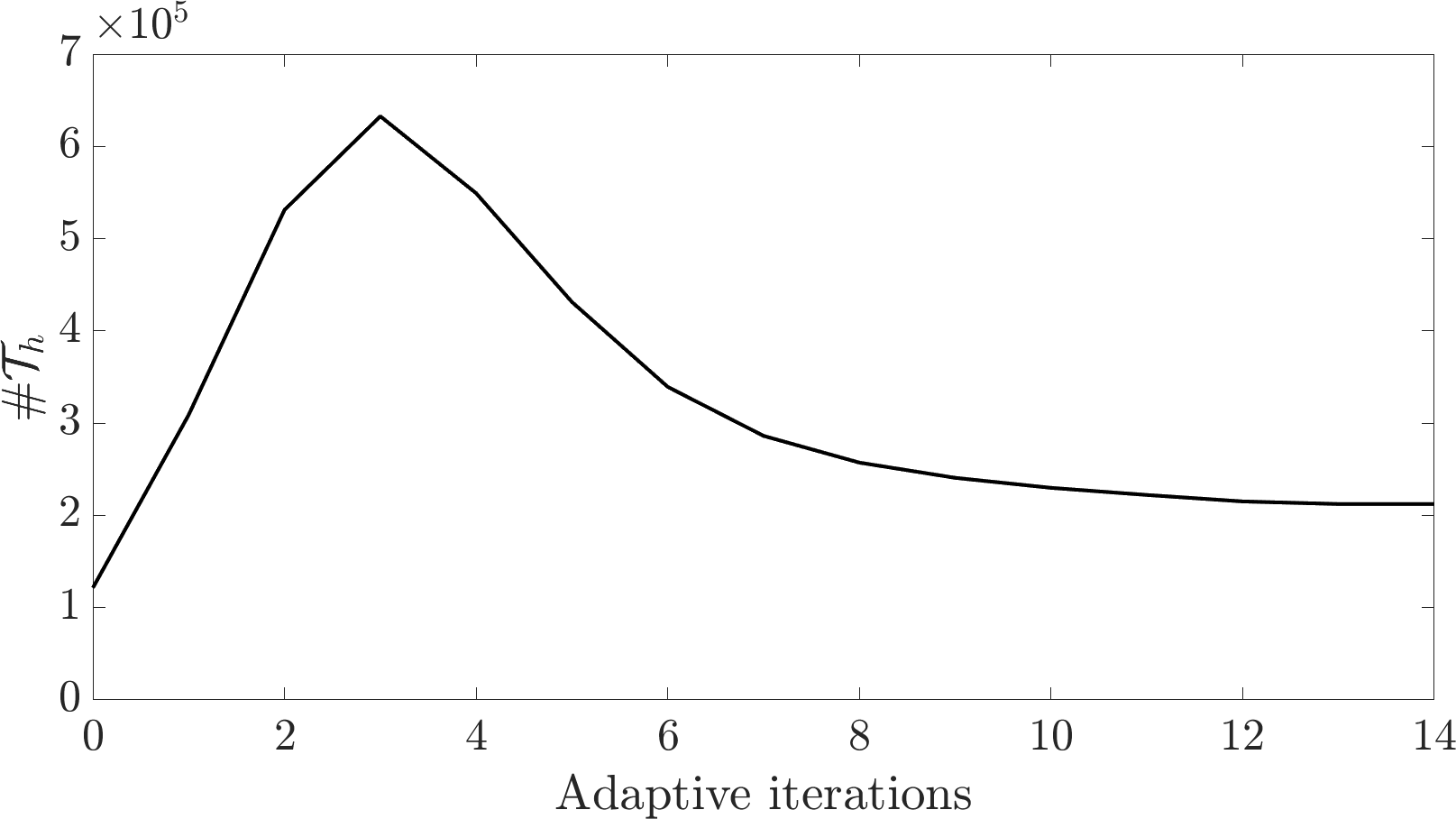}
\caption{Convergence history of LEVITY: evolution of the mesh cardinality: bridge (left); jet engine bracket (right).}\label{fig:bridge_convergence}
\end{center}
\end{figure}

The left panel in Figure~\ref{fig:bridge_convergence} shows the trend of the mesh cardinality throughout the $17$ mesh adaptations. The adoption of a coarse initial mesh justifies the abrupt increment of the number of mesh elements. However, six mesh adaptations suffice to reach a stationary regime in the mesh evolution.
\\
Finally, it is interesting to remark that the adaptation phase is computationally cheaper with respect to the evolution of the level set, taking only $15.26\%$ of the overall runtime\footnote{The simulations are run on a computer with 8 GB of RAM, a CPU with 6 i7-3930K cores and a maximum frequency of 3.20 GHz.}.

\subsection{The jet engine bracket}\label{sec:bracket}

This test case is inspired by the well-known challenge promoted by GE (General Electric) and GrabCAD in 2013\footnote{https://grabcad.com/challenges/ge-jet-engine-bracket-challenge}. The optimization goal merges performance requirements for the stiffness with a reduction of the structure weight, consistently with problem \eqref{level_set}.

\begin{figure}[H]
\begin{center}
\includegraphics[width=0.3\columnwidth]{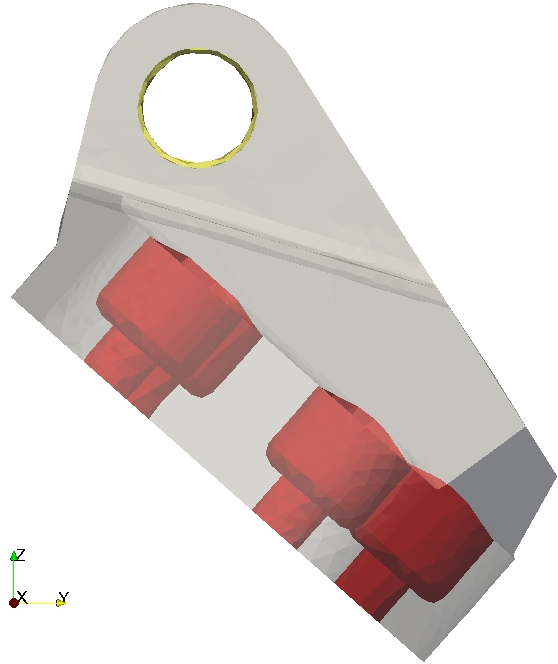}\qquad
\includegraphics[width=0.5\columnwidth]{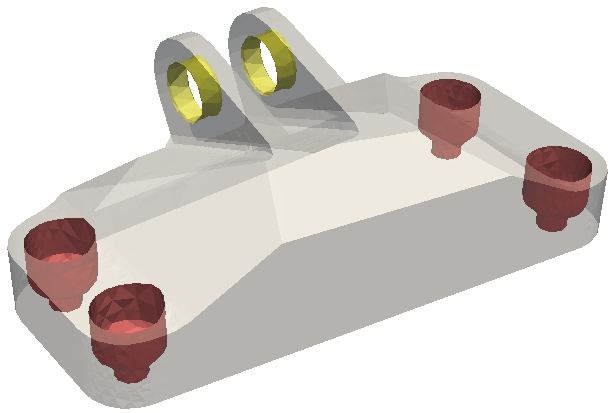}
\caption{The jet engine bracket. Design domain with color-coded $\Gamma_D$ (red) and $\Gamma_t$ (yellow) boundary portions: lateral (left) and front (right) views.}\label{fig:bracket_domain}
\end{center}
\end{figure}

The design domain $\Omega$ is illustrated in Figure~\ref{fig:bracket_domain}. A load ${\bf t}=(0, 0, 1$e-$03)^T$ is applied to the boundary portion 
$\Gamma_t$, which is yellow-highlighted in the right panel. Homogeneous Dirichlet boundary conditions for the displacement are imposed on the boundary portion $\Gamma_D$ coinciding with the four bolt holes highlighted in red. Since it is evident that these holes have to be preserved during the design process, we exclude these areas from the optimization. 
Finally, as far as the adopted material is concerned, we select $E=1$ and $\nu=0.3$.

The topology optimization performed by LEVITY is identified by the following choice for the input parameters: ${\tt CTOL} = 1$e-$04$, ${\tt TOL}=65$, ${\tt ATOL}=5$e-$03$, ${\tt kmax}=700$, ${\tt kStart}=200$, ${\tt kAdapt}=25$, $\varphi_h^0=1$, $\T_h^{\tt 0}$ a uniform mesh with $121472$ tetrahedra, ${\tt grade}=1$, $h_{iso}=5$, $\Delta t=0.1$, $\alpha=0.3$, $V_0 = 463368$, $\chi_{min}=1$e-$03$, $\tau=1.5$, and $\beta=20$.

\begin{figure}[H]
\begin{center}
\includegraphics[width=0.45\columnwidth]{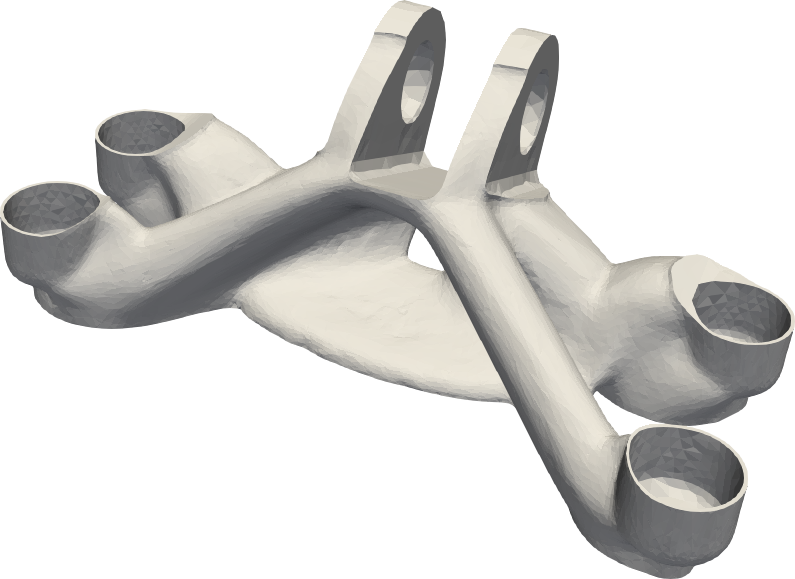}
\includegraphics[width=0.45\columnwidth]{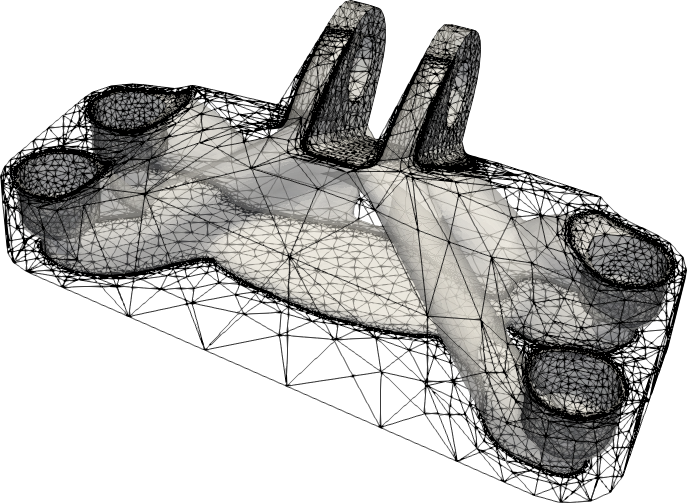}
\includegraphics[width=0.45\columnwidth]{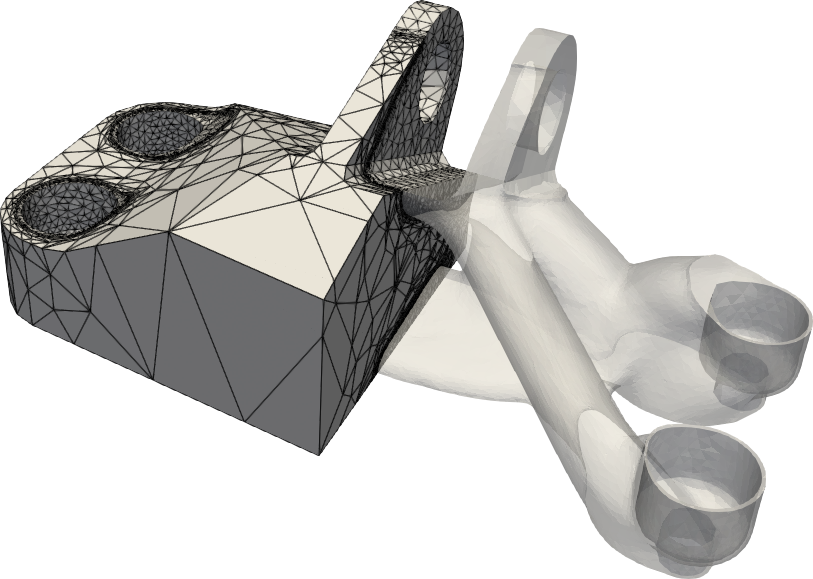}
\includegraphics[width=0.45\columnwidth]{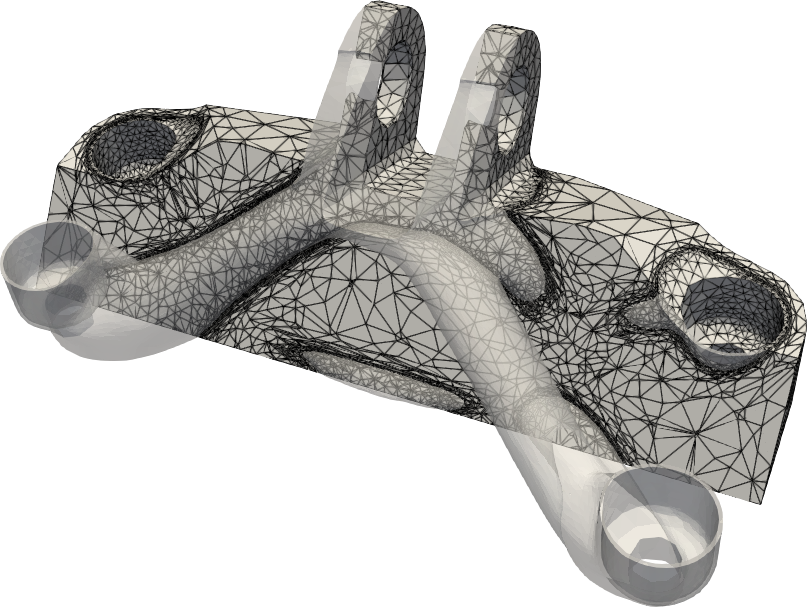}
\caption{The jet engine bracket. Output of LEVITY algorithm: final layout $\overline{\Sigma}_h$ (top-left) and graded mesh superimposed to the optimized structure (top-right); cut views of the graded meshes of $\Omega$ superimposed to the optimized layout (bottom).}\label{fig:bracket_levelset}
\end{center}
\end{figure}

The convergence is reached after $525$ iterations, with $14$ mesh adaptations.
The returned final layout is characterized by a compliance equal to $0.21$ and is shown in Figure~\ref{fig:bracket_levelset} (top-left). The adapted graded mesh presents highly anisotropic elements along the structure contour, isotropic tetrahedra in the internal part, and a coarse tessellation of the design domain portion $\Omega \setminus \overline{\Sigma}_h$, as highlighted in Figure~\ref{fig:bracket_levelset} (top-right, bottom).
\\
Similarly to the bridge case study, the anisotropic grid adaptation procedure quickly reaches a stagnation on the mesh cardinality (see the right panel in Figure~\ref{fig:bridge_convergence}), while requiring a minimal percentage ($15.23\%$) of the whole computational effort. The number of elements associated with the final graded mesh is very limited ($212360$ tetrahedra). We do expect that Algorithm~\ref{level_set_algo} does not deliver an optimized structure with the same smoothness for a uniform mesh $\mathcal{T}_h^{\tt 0}$ characterized by a similar cardinality.

\begin{figure}[H]
\begin{center}
\includegraphics[width=0.45\columnwidth]{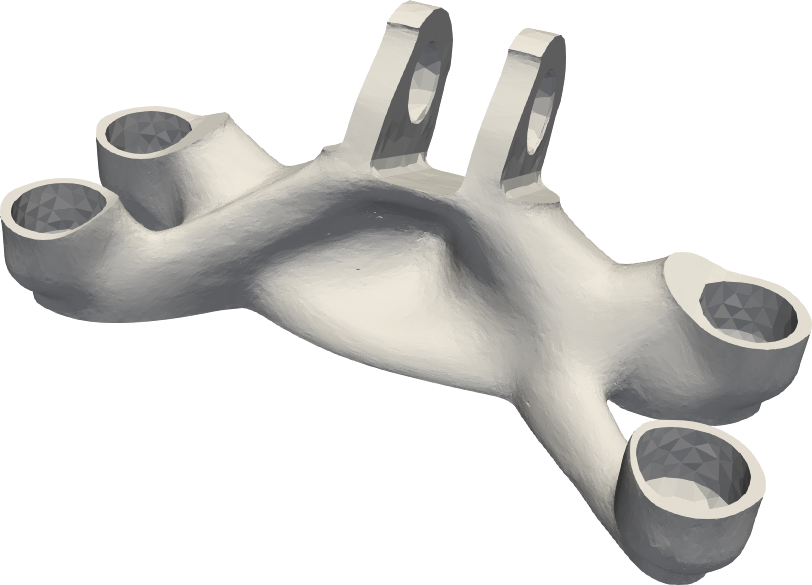}
\includegraphics[width=0.45\columnwidth]{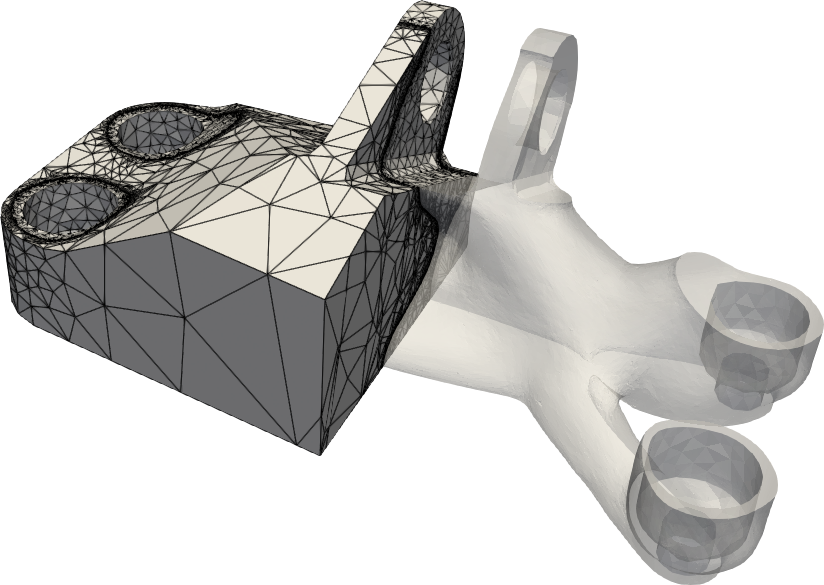}
\caption{The jet engine bracket. Output of the SIMPATY$_G$ algorithm: final layout $\overline{\Sigma}_h$ (left) and cut view of the graded mesh of $\Omega$ superimposed to the optimized layout(right).}\label{fig:bracket_simp}
\end{center}
\end{figure}

As a last check, we investigate the outcome provided by the SIMPATY$_G$ method for $\gamma=5$e-$03$ in \eqref{LG}. The optimized bracket is shown in Figure~\ref{fig:bracket_simp} together with a cut view of the surface graded mesh. The compliance associated with this layout is equal to $0.25$ and the mesh cardinality is $205266$.

A cross-comparison between Figures~\ref{fig:bracket_levelset} and~\ref{fig:bracket_simp} highlights that the load path towards the bolt holes is essentially the same, thus identifying four major branches. On the contrary, the topology of the central body is different, with the presence of an additional hole in the LEVITY configuration.
\\
The smoothness of the jet engine bracket surface is fully comparable in the two cases, as well as the limited computational time characterizing the adaptation step ($15.23\%$ and $5.12\%$ of the whole time for LEVITY and SIMPATY$_G$, respectively).

\section{Conclusions}

This paper introduces a new algorithm named LEVITY (LEVel set with mesh adaptivITY) for topology optimization, which leads to outperform a standard level set approach in terms of computational efficiency. The verification carried out in a $2$D setting shows that LEVITY essentially preserves all the good properties of the level set formulation, such as the independence of the initial topology and mesh, the smoothness of the layout, and the involvement of few parameters in the design process. In this context, the most relevant advantage yielded by an anisotropic adapted mesh is the smoothness guaranteed also to the characteristic function $\chi_{\varphi,h}$ (compare Figures~\ref{fig:fixed_clc} and \ref{fig:aniso_clc}). Moreover, the graded feature of the computational mesh ensures the reliability of the final design in terms of mechanical performances.

LEVITY algorithm shows the full potentiality due to the introduction of anisotropic adapted grids in Section~\ref{sec:3d}, where realistic configurations are optimized. In particular, the computational module {\tt adaptMesh} leads to final layouts exhibiting a very smooth contour by demanding an almost negligible computational effort (around $15 \%$ of the whole CPU time). This process is fully automatic and very few iterations suffice to guarantee the mesh cardinality stagnation (see Figure~\ref{fig:bridge_convergence}).

In Sections~\ref{numerical_sec} and~\ref{sec:3d}, we take into account another topology optimization process, based on anisotropic mesh adaptation, namely SIMPATY (SIMP with mesh AdaptivITY) algorithm. 
Although beyond the purpose of this paper, a comparison between LEVITY and SIMPATY leads to some considerations. SIMPATY suffers from the dependence of the optimized structure on the initial topology in contrast to LEVITY (compare Figures \ref{fig:levels_CLSC} and~\ref{fig:simp_sens}). The regularized compliance in \eqref{LG} allows us to get rid of this limitation (as confirmed by Figure~\ref{fig:levels_simp}) and to make the two methodologies more consistent, as highlighted by the similar (although not identical) returned topologies (compare Figure~\ref{fig:levels_CLSC} with~\ref{fig:levels_simp} and Figure~\ref{fig:bracket_levelset} with~\ref{fig:bracket_simp}). The optimization step characterizing SIMPATY unavoidably involves some parameters to be tuned that may depend on the specific application. In this regard, the diffusive process in \eqref{evolution} turns out to be more straightforward, since $\tau$ is the only parameter to be properly selected. This feature becomes a drawback when the user is interested in a fine-tuning of the design procedure. Additionally, the computation of the topological derivative $d_t \overline{F}$ in \eqref{evolution} may become an issue for optimization settings more complex with respect to the one in \eqref{level_set}.
Finally, we remark that the employment of a graded grid both in LEVITY and SIMPATY procedures make the two algorithms fully comparable from a mechanical viewpoint, as shown by the very similar values for the compliance (see Tables~\ref{tab_levsens} and \ref{tab_simpg}).
\\
To sum up, we observe that the addition of a mesh adaptation routine into a standard SIMP or level set topology optimization approach is instrumental to set an efficient design tool capable of striking a balance between reliability and computational affordability, and of delivering free-form structures. The main discrepancies between SIMP and the level set are essentially preserved by SIMPATY and LEVITY.

Concerning future developments of this research, we plan
to tackle other optimization frameworks characterized by various objectives and constraints by exploiting the full generality of the proposed optimal design procedure. As a second perspective, we aim at optimizing the implementation of Algorithm~\ref{lev_a} for an efficient management of the different blocks, also employing parallel computing architectures. Moreover, we also foresee the introduction of automatic differentiation in the algorithm pipeline to easily assemble the sensitivities for diverse application contexts.

\section*{Acknowledgments}
This research is part of the activity of the METAMatLab at Politecnico di Milano.
\\
SP and SM thank the PRIN research grant n.20204LN5N5 \textit{Advanced Polyhedral Discretisations of Heterogeneous PDEs for Multiphysics Problems} and the INdAM-GNCS 2022 Project \textit{Metodi di riduzione computazionale per le scienze applicate: focus su sistemi complessi}. 



\bibliographystyle{elsarticle-num} 
\bibliography{mybib}

\end{document}